\documentclass[aps,pra,twocolumn,notitlepage,superscriptaddress,showpacs,10pt,longbibliography]{revtex4-2} 

\usepackage{graphicx}
\usepackage{dcolumn}
\usepackage{bm}
\usepackage{amsmath}
\usepackage{amssymb}
\usepackage{amstext}
\usepackage{amsthm}
\usepackage{txfonts}
\usepackage{pxfonts}
\usepackage[table]{xcolor}
\usepackage[colorlinks=true, allcolors=blue]{hyperref}
\usepackage{color}
\usepackage{physics}
\usepackage{xcolor}
\usepackage[utf8]{inputenc}
\usepackage[T1]{fontenc}
\usepackage{mathptmx}
\usepackage{etoolbox}
\usepackage{appendix}
\usepackage{svg}

\makeatletter
\let\openbox\@undefined
\makeatother

\begin{document}

\preprint{APS/123-QED}

\title{Observation of mechanical kink control and generation via phonons}
 
\author{Kai Qian}
\affiliation{Department of Mechanical and Aerospace Engineering, University of California San Diego, La Jolla, CA 92093, USA}

\author{Nan Cheng}
\thanks{These two authors contributed equally.}
\affiliation{Department of Physics, University of Michigan, Ann Arbor, MI 48109, USA}

\author{Francesco Serafin}
\thanks{These two authors contributed equally.}
\affiliation{Department of Physics, University of Michigan, Ann Arbor, MI 48109, USA}
\affiliation{
Department of Physics and Materials Science, University of Luxembourg, 2 Avenue de l'Universite L, 4365 Esch-sur-Alzette, Luxembourg}

\author{Kai Sun}
\affiliation{Department of Physics, University of Michigan, Ann Arbor, MI 48109, USA}

\author{Georgios Theocharis\footnotemark[1]}
\thanks{These three authors are co-corresponding authors.
Emails: georgiostheocharis@gmail.com, maox@umich.edu, nboechler@ucsd.edu}
\affiliation{Laboratoire d’Acoustique de l’Université du Mans (LAUM), UMR 6613, Institut d’Acoustique - Graduate School (IA-GS), CNRS, Le Mans Université, France}

\author{Xiaoming Mao\footnotemark[1]}
\thanks{These three authors are co-corresponding authors.
Emails: georgiostheocharis@gmail.com, maox@umich.edu, nboechler@ucsd.edu}
\affiliation{Department of Physics, University of Michigan, Ann Arbor, MI 48109, USA}

\author{Nicholas Boechler\footnotemark[1]}
\thanks{These three authors are co-corresponding authors.
Emails: georgiostheocharis@gmail.com, maox@umich.edu, nboechler@ucsd.edu}
\affiliation{Department of Mechanical and Aerospace Engineering, University of California San Diego, La Jolla, CA 92093, USA}
\affiliation{Program in Materials Science and Engineering, University of California San Diego, La Jolla, CA 92093, USA}

\begin{abstract}

Kinks (or domain walls) are localized transitions between distinct ground states associated with a topological invariant, and are central to many phenomena across physics, from condensed matter to cosmology. While phonon (\textit{i.e.}, small-amplitude vibration) wave packets have been theorized to deterministically interact with kinks and initiate their movement, this interaction has remained elusive in experiments, where only uncontrollable stochastic kink motion generated by thermal phonons or dislocation glide by low-frequency quasi-static loading have been observed. This is partly because all physical systems that support kinks are, at some level, discrete, making deterministic phonon control of kinks extremely challenging due to the existence of Peierls-Nabarro (PN) barrier. Here, we demonstrate, for the first time, experimental observation of phonon-mediated control and generation of mechanical kinks, which we enable using a topological metamaterial that constitutes an elastic realization of the Kane-Lubensky chain model. Our metamaterial overcomes the PN barrier by supporting a single, topologically protected kink that requires zero energy to form and move. Using simulations that show close agreement with our experimental observations, we also reveal unique dynamics of phonon interplay with highly discrete kinks, including long-duration motion and a continuous family of internal modes, features absent in other discrete nonlinear systems. This work introduces a new paradigm for topological kink control, with potential applications in material stiffness tuning, shape morphing, locomotion, and robust signal transmission.

\end{abstract}

\maketitle

\section{Introduction}

Kinks (or domain walls) are localized transitions between distinct ground states that are associated with a topological invariant, and are of great significance in physics, across fields ranging from condensed matter to cosmology~\cite{dauxois2006physics, vachaspati2007kinks}. While precise definitions vary depending on the field of research \cite{chaikin1995principles}, given the one-dimensional (1D) setting considered herein, we will use ``kinks'' and ``domain walls'' interchangeably. A particular subset of kinks is what we define as ``mechanical kinks'', wherein the transition between ground states involves a gradient of displacement (\textit{i.e.}, strain). Several settings that see mechanical kinks, spanning a wide range of length and time scales, are crystal plasticity~\cite{frenkel1939theory,weiner1964dislocation,atkinson1965motion,earmme1974breakdown,flytzanis1977solitonlike}, formation and propagation of denaturation bubbles in DNA~\cite{komarova2005nonlinear}, interfaces in polyacetylene~\cite{su1979solitons}, ferroelectric domains~\cite{bruce1981scattering,nataf2020domain}, colloidal self-assembly \cite{menath2023acoustic,he2024regulating}, as well as bistable and topological states in mechanical metamaterials~\cite{chen2014nonlinear,zhou2017kink,mao2018maxwell,zhang2019programmable,deng2020characterization,upadhyaya2020nuts,lo2021topology,zhou2021amplitude,sun2021fractional,deng2022topological,nunez2023fractional,zhou2024static}. Given their ubiquity and their impact on mechanical and transport properties of materials, understanding how mechanical kinks can be controlled and generated is a topic of significant interest.

Many studies have experimentally shown the ability of phonons (or small-amplitude vibrations) to initiate and affect the movement of mechanical kinks. However, in each of these studies, the vibration is either so far below the characteristic vibrational frequencies of the system that it can be considered a quasi-static loading \cite{gorman1969mobility,gremaud1987coupling}, or the phonons are applied in such a spatially extended manner that the resulting kink motion is stochastic (\textit{e.g.}, acoustic annealing in colloidal crystals \cite{menath2023acoustic,he2024regulating}, or thermally-induced motion of dislocations in metals \cite{conrad1964thermally, hutchison1965thermally} and kinks in polyacetylene \cite{ogata1986brownian})~\footnote{Another, albeit less, related area is phonon interplay with multistable systems supporting transition waves (\textit{e.g.}, Ref.~\cite{khomeriki2008tristability, wu2018metastable, hwang2021extreme}). We note that these systems do not support kinks as they do not have symmetric ground states and use high-amplitude excitation to initiate transitions}. As a result, we ask: is it possible to \textit{control} and \textit{generate} mechanical kinks with phonons in a deterministic manner? 

Theoretical and computational studies have given further insight to this question, exploring the interaction between phonon wave packets and mechanical kinks via canonical models such as the $\phi^4$ \cite{hasenfratz1977interaction, wada1978brownian,theodorakopoulos1979dynamics, theodorakopoulos1980lattice, klein1980kink, ishiuchi1980brownian,ogata1984momentum, abdelhady2011wave} and sine-Gordon (sG) systems~\cite{theodorakopoulos1980dynamics,wada1982brownian}. One challenge in precisely controlling kinks via phonons, as revealed by studies of such models, stems from the fact that all the aforementioned physical systems supporting kinks are, at some level, discrete, 
where the width of the kinks is comparable with the lattice spacing \cite{frenkel1939theory,weiner1964dislocation,atkinson1965motion,earmme1974breakdown,flytzanis1977solitonlike,su1979solitons,bruce1981scattering,nataf2020domain,komarova2005nonlinear,chen2014nonlinear,zhou2017kink,mao2018maxwell,zhang2019programmable,deng2020characterization,upadhyaya2020nuts,lo2021topology,zhou2021amplitude,deng2022topological,zhou2024static,menath2023acoustic,he2024regulating, sun2021fractional,nunez2023fractional}. This is important because of the emergence of the static Peierls-Nabarro (PN) potential barrier in discrete settings \cite{peierls1940size, nabarro1947dislocations, kivshar1993peierls,braun2004frenkel,chirilus2014sine,kevrekidis2019dynamical}, which arises from the breaking of the translational invariance present in continuum systems \cite{dauxois2006physics,chirilus2014sine,kevrekidis2019dynamical,dauxois2006physics,chirilus2014sine}. The PN barrier restricts the mobility of the kinks---typically causing moving kinks to lose energy through phonon radiation and eventually become pinned~\cite{peyrard1984kink}. Although studies have identified discrete kink models absent of a PN barrier (\textit{i.e.}, barrier-free) arising from exceptional discretizations \cite{speight1994kink,speight1997discrete,speight1999topological,flach1999moving,savin2000moving,kevrekidis2003class,aigner2003new,cooper2005exact, dmitriev2005discrete,oxtoby2005travelling,barashenkov2005translationally,dmitriev2006standard,dmitriev2006exact,kevrekidis2019dynamical,saadatmand2024phonons}, these cases are primarily of mathematical interest, with no known physical realization. 
Beyond the challenge of controlling kinks, their generation is also difficult, as it requires initiation either through large-amplitude excitations \cite{scharf1992sine,mohammadi2021kink,simas2024generation} or within active systems \cite{woodhouse2018autonomous, ghatak2020observation, veenstra2024non}.

\begin{figure}[ht!]
    \centering
    \includegraphics[width=8.6cm]{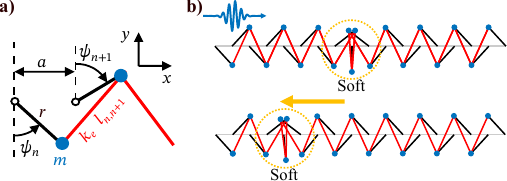}
    \caption{\textbf{Overview of a phonon wave packet moving the static zero-energy kink in the KL chain.}
    (a) Schematic of a chain section, where black lines represent massless rotors with radius $r$ and lattice constant $a$, blue circles represent point masses $m$, and red lines represent linear normal springs with spring constant $k_e$ and instantaneous length $l_{n,n+1}$ with $n$ and $n+1$ as rotor indices.
    (b) Kink configuration before (top) and after (bottom) phonon wave packet interaction. 
    The yellow dashed circle indicates the center of the zero-energy kink (\textit{i.e.}, soft). 
    }
    \label{fig:overview}
\end{figure}

In this paper, we report the first experimental observation of mechanical kink control and generation via phonons. 
In particular, the kinks we control and generate are highly discrete (width less than two lattice spacings), which is important, as the strength of the PN barrier typically increases as the kink width decreases \cite{hobart1965peierls,nabarro1989peierls,braun2004frenkel}.
To achieve this, we create an elastically coupled realization of the Kane-Lubensky (KL) chain model~\cite{kane2014topological}---a 1D topological metamaterial---as shown in Fig.~\ref{fig:overview}(a,b). We note that, in contrast to the other aforementioned kink-supporting systems, mechanical metamaterials further facilitate experimental observation of such control, as they are typically macroscopic with significantly lower characteristic frequencies \cite{chen2014nonlinear,zhang2019programmable,deng2020characterization,zhou2021amplitude,deng2022topological,nunez2023fractional,zhou2024static}. The special feature of the KL chain, as regards kink control via phonons, is that it supports a single topologically protected kink that requires zero energy to move quasi-statically (also known as a nonlinear zero mode or a mechanism), resulting in a zero PN barrier \cite{chen2014nonlinear,zhou2017kink, mao2018maxwell,upadhyaya2020nuts,lo2021topology}. Throughout this paper, we refer to this kink as a ``zero-energy kink.''

In addition to experimental observations, we provide a comprehensive numerical study of the phonon spectrum of the KL chain featuring the zero-energy kink.
First, we find the existence of several ``internal'' modes  \cite{kevrekidis2000dynamics,dauxois2006physics,chirilus2014sine,kevrekidis2019dynamical}. Internal modes have been shown to play a crucial role in kink dynamics, as they can store and release energy, leading to resonance effects during kink collisions \cite{peyrard1983kink} or interactions with inhomogeneities \cite{kivshar1991resonant,fei1992resonant}. Experimental evidence of internal modes has been reported in a variety of systems, from polyacetylene \cite{vardeny1986detection} to crystals of trapped ions \cite{mielenz2013trapping,brox2017spectroscopy}, underscoring their broad relevance in different physical contexts. In our KL chains, we identify a continuously varying set of finite-frequency internal modes associated with smoothly varying kink states. This continuity is distinct from other nonlinear discrete systems, which typically support only a finite set of kink solutions with corresponding internal modes, such as those associated with onsite- or intersite-centered static kink solutions \cite{dauxois2006physics,chirilus2014sine,kevrekidis2019dynamical}. 

Besides the phonon spectra of the zero-energy kink, we also numerically reveal two other distinctive features of phonon-kink interaction in the KL chain. First, we observe long-duration kink motion with no apparent slowdown after interaction with a phonon wave packet. This contrasts with the behavior seen in the discrete $\phi^4$ chain, where the kink eventually slows down and stops due to phonon radiation induced by the PN barrier following wave packet interaction. The long-lasting motion observed here has important implications for kink control applications, as it suggests that the kinks could be driven further with less energy expenditure. Second, we observe a range of phonon-kink interactions depending on the geometric properties of the chain, including both kink attraction and repulsion, as well as varying interactions depending on the amplitude and frequency of the phonon wave packet and the kink's center position, constituting additional knobs for phonon-mediated kink control. 

Building upon this new capacity for kink control, we highlight here additional application-relevant benefits from these phenomena existing in the KL chain, as well as potential future physical manifestations. Because the KL chain is a type of topological metamaterial, and the kink is a topologically protected defect, one can imagine future uses in robust signal transmission \cite{ni2023topological}. Similarly, because the kink in the KL chain represents a transition between states of, ideally, localized zero and infinite stiffness \cite{rocklin2017transformable,mao2018maxwell}, one can imagine remote control of extreme material stiffness, \textit{e.g.}, soft on one side and stiff on the other, \textit{vice versa}, or stiff on both. Further functionalities that have seen benefit from topological kink movement include locomotion \cite{deng2022topological} and shape-shifting materials \cite{rodriguez2023mechanical}, 
both of which would benefit from low-energy control and long-distance kink transport. Our newly discovered phenomena may also inspire the search for analogous nanoscale or molecular-level mechanisms~\cite{peplow2015tiniest}. Potential analogies may be drawn with, \textit{e.g.}, rotor-like behaviors in bacterial flagellar motors~\cite{nirody2017biophysicist,singh2024cryoem}, DNA~\cite{shi2022sustained}, or nanoelectromechanical systems~\cite{kim2014ultrahigh}. 

\section{Model}

We start by reviewing the key features of the KL chain and its kink states (\textit{i.e.}, kink-containing configurations) to formulate the problem.
We describe the rotor angles in the chain with $\psi_n$ (where $n$ is the rotor index), measured clockwise for even rotors and counterclockwise for odd ones, with the rotor radius, lattice constant, spring stiffness, and instantaneous length denoted by $r$, $a$, $k_e$, and $l_{n,n+1}$, respectively, as shown in Fig.~\ref{fig:overview}(a). 
The KL chain exhibits two homogeneous configurations when $\psi_n = \pm \bar{\psi}$ ($\bar{\psi} \in [0, \pi/2]$), referred to as the right- (RPS) and left-polarized state (LPS), respectively, resulting in a uniform unstretched spring length $l_{n,n+1}=\bar{l}$ throughout the chain~\cite{kane2014topological,chen2014nonlinear,mao2018maxwell}. 
In the linear regime, these configurations give rise to a topological polarization vector $R_T$, leading to an exponentially localized zero-energy mode at the edge pointed to by $R_T$~\cite{kane2014topological,chen2014nonlinear,mao2018maxwell}.
This localized zero mode creates a soft edge, leaving the rest of the chain rigid.

When the zero mode in the homogeneous chain is quasi-statically driven to the nonlinear regime, it smoothly deforms into a kink state that separates regions of opposite polarizations and exhibits zero energy \cite{chen2014nonlinear}, \textit{i.e.}, the kink is ``soft,'' as marked in Fig.~\ref{fig:overview}(b).
This kink exhibits two distinct continuum limits---one can be described by a field theory resembling $\phi^4$ \cite{dauxois2006physics,chen2014nonlinear,manton2004topological,kevrekidis2019dynamical} and the other by the sine-Gordon (sG) field theory \cite{dauxois2006physics,chen2014nonlinear,chirilus2014sine}---which depend on the geometry of the unit cell \cite{chen2014nonlinear}. 
The width of the kink is denoted by $w_0=2\kappa$, where $\kappa=-a/\ln{\abs{{(d-1)}/{(d+1)}}}$ is the penetration depth of the zero mode in the homogeneous chain, with $d=2r\sin{\bar{\psi}}/a$ a dimensionless geometric index \cite{chen2014nonlinear}. When $d\ll 1$, $\kappa\approx {a}/{(2d)}$, diverging as $d\to 0$, and the kink can be described by a continuum theory similar to $\phi^4$ \cite{chen2014nonlinear}. When $d\gg 1$, $\kappa\approx ad/2$, diverging as $d\to \infty$, and the kink can be described by the continuum sG theory, where the representation involves only alternating rotors (\textit{i.e.}, all odd or even rotors) \cite{chen2014nonlinear}. 
The feature of possessing two continuum limits for the kink in the KL chain is distinct from kinks in other systems, such as a discrete $\phi^4$ chain \cite{dauxois2006physics}, which exhibits a single continuum limit when the wavelength is much longer than the length of the unit cell. 

\begin{figure*}[ht!]
    \centering
    \includegraphics[width=17.2cm]{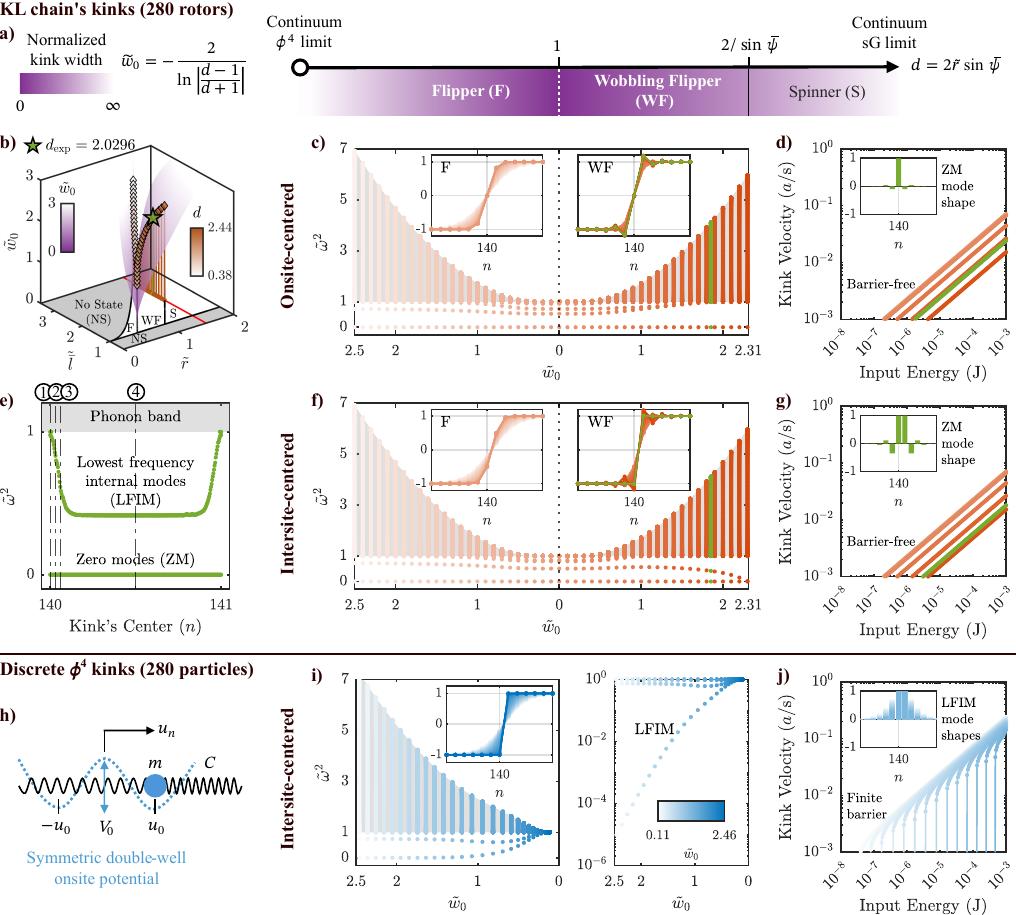}
    \caption{
    \textbf{Comparison of shapes, phonon spectra, and short-time dynamics of KL chain's F/WF-phase kinks (a-g) and discrete $\phi^4$ kinks (h-j).}
    (a) Phase and kink width diagram of the zero-energy kink of the KL chain as a function of the dimensionless index $d$ (introduced in Ref.~\cite{chen2014nonlinear}), where $\tilde{r} \equiv r/a$ is the normalized radius of rotors. 
    (b) Geometric phase diagram of the zero-energy kink in terms of $\tilde{r}$ and normalized unstretched spring length $\tilde{\bar{l}} \equiv l/a$, where the red line indicates $\tilde{r} = 1.5$, orange diamonds indicate kink states shown in (c,d,f,g), and the green star denotes the kink state examined in experiments. Green is further used in (c-g), as well as Figs.~\ref{fig:longtime_phonon-kink_dynamics_1} and \ref{fig:longtime_phonon-kink_dynamics_2}, to denote the chains with the same unit cell geometry as in experiments.
    (c,f,i) Normalized eigenfrequencies of kinks as a function of kink widths, where (c) and (f,i) show onsite- and intersite-centered solutions, respectively.  
    Shaded area indicates the phonon band of the infinite homogeneous chain. 
    Insets show zoomed-in, normalized static kink solutions for KL chains ($-\sin{\psi_n}/\sin{\bar{\psi}}$ \cite{chen2014nonlinear}) and discrete $\phi^4$ chains ($u_n/u_0$).
    For the F- and WF-phase kinks of the KL chain, the lowest eigenmode is always zero, suggesting ``barrier-free'' kink propagation regardless of kink width (or equivalently, kink discretization).
    (e) In-gap modes of the experimental kink state as a function of kink's center position.
    \textcircled{1} and \textcircled{4} represent onsite- and intersite-centered kinks, respectively, while
    \textcircled{2} and \textcircled{3} correspond to intermediate kink states between them. 
    (d,g,j) Short-time kink propagation velocity versus initial input energy for kink states shown in (c,f,i), where the $y$-axis is truncated to $10^{-3}$ due to the velocity resolution.
    The inset in (d,g) shows zoomed-in, normalized initial velocities for rotors in the KL chain with experimental unit cell geometry, whose magnitudes are based on the mode shapes of corresponding zero mode.
    The inset in (j) shows zoomed-in, normalized initial velocities for particles in the discrete $\phi^4$ chain, whose magnitudes are based on the mode shapes of the lowest frequency internal mode (LFIM) for each chain configuration.
    In (d,g), only cases with $\tilde{w}_0=0,0.5,1,1.5,2$, along with the experimental case, are presented. In (j), only cases where $\tilde{w}_0>1$ are shown.
    (h) Schematic of a discrete $\phi^4$ chain  section. 
    } 
    \label{fig:kink_dynamics}
\end{figure*}

Between these two continuum limits, the kink state can be classified into three distinct phases \cite{chen2014nonlinear} based on the range of rotor angle changes during kink propagation, 
characterized by $d$: 
i) the ``Flipper'' (F), where $d<1$ and ${\psi_n}$ varies between $-\bar{\psi}$ and $+\bar{\psi}$ during kink propagation; 
ii) the ``Wobbling Flipper'' (WF), where $1<d<2/\sin{\bar{\psi}}$ and the rotors move beyond their homogeneous positions but cannot complete full rotations; and 
iii) the ``Spinner'' (S), where the rotors can complete full rotations during kink propagation. We note that the previously introduced continuum limits occur deep within the F and S phases, respectively, as illustrated in Fig.~\ref{fig:kink_dynamics}(a). 
In Fig.~\ref{fig:kink_dynamics}(a), we illustrate the dependence of normalized kink widths $\tilde{w}_0 \equiv {w}_0/a$ on $d$ and highlight the difference between the F/WF- and S-phase kinks by a solid line, which indicates the topological change in the configuration space of the KL chain in the linkage limit \cite{chen2014nonlinear}.
In this paper, we investigate F- and WF-phase kinks of the KL chain, and leave the investigation of S-phase kinks for future work.
Noting that the unit cell geometry that results in a given phase $d$ is non-unique (see the SM for details), in Fig.~\ref{fig:kink_dynamics}(b), we also plot the kink width for F- and WF-phase kinks in terms of two dimensionless geometric parameters, $\tilde{\bar{l}} \equiv \bar{l}/a$ and $\tilde{r} \equiv r/a$. 

\section{Results}

\subsection{Computed results of existence, stability and short-time dynamics of zero-energy kinks}

In this section, we computationally study the existence, stability and short-time dynamics of F- and WF-phase kinks in the KL chain with varying widths (while keeping the lattice spacing $a$ fixed) and compare them with those of discrete $\phi^4$ kinks. As mentioned earlier, deep into the F phase ($d\ll1$), the KL chain can be described by a continuum version of the $\phi^4$ model. 
To highlight the unique properties of the KL chain---specifically, the barrier-free propagation and stability of its zero-energy kink---we compare it with the discrete $\phi^4$ chain \cite{dauxois2006physics}, a prototypical lattice system extensively studied over the past decades. In this section, we quantify their similarities and their key differences.
For our discrete $\phi^4$ chains, we consider a 1D mass-spring chain with a symmetric double-well onsite potential~\cite{dauxois2006physics}, as shown in Fig.~\ref{fig:kink_dynamics}(h), where $m$ is the particle mass, $C$ the coupling constant, $V_0$ the potential energy well depth, $u_n$ the particle displacement, and the minima of the potential are located at $u_n=\pm u_0$ (see the SM for further details on the equations of motion).
We summarize our results in Fig.~\ref{fig:kink_dynamics}, 
where (c,d) and (f,g) show onsite- and intersite-centered kink results in KL chains, respectively, and (i,j) show intersite-centered $\phi^4$ kink results.
Both the KL chain and the discrete $\phi^4$ chain consist of $280$ sites (\textit{i.e.}, rotors or particles) with open boundary conditions. Their static kink solutions and their stability are numerically calculated using the Newton-Raphson method \cite{gilat2013numerical}, based on their respective equations of motion, and the corresponding Jacobian matrix, except for results in Fig.~\ref{fig:kink_dynamics}(e) (see the SM for details). 

To vary the kink widths: in the KL chain, we fix $r$ and vary $l$ (equivalent to varying $\Bar{\psi}$ in the homogeneous chain); while in the discrete $\phi^4$ chain, we vary $C$. The kink width of discrete $\phi^4$ kinks is defined as $\tilde{w}_0 = 1/\beta$, where $\beta$ is obtained by fitting $u_n = \alpha\tanh{\left(\beta(n-\gamma)\right)}$.
Insets in Fig.~\ref{fig:kink_dynamics}(c,f) and (i) show the stable onsite- and intersite-centered F/WF-phase kinks in the KL chain, and intersite-centered $\phi^4$ kinks, respectively.
The KL chain kinks shown in Fig.~\ref{fig:kink_dynamics}(c,d,f,g) also correspond to the red line and the rhombus markers in Fig.~\ref{fig:kink_dynamics}(b). We note that, although the onsite-centered kink solutions are unstable in discrete $\phi^4$ chains, they remain stable for F/WF-phase kinks in the KL chain, showing that the stable kinks can exist anywhere along the KL chain. See the SM for unstable onsite-centered kink solutions and their spectra in discrete $\phi^4$ chains.

Figure~\ref{fig:kink_dynamics}(c,f) and (i) show the eigenfrequencies of the finite chains for the kink states shown in the insets for the KL and discrete $\phi^4$ chains, respectively, as a function of kink width. 
Shaded areas denote the phonon band ranges of the infinite homogeneous configurations for both chains.
Each frequency is normalized by the lower limit of their respective phonon band.
Comparing Fig.~\ref{fig:kink_dynamics}(c,f) and (i), we first observe that the phonon band collapses at $\tilde{w}_0=0$, during the transition between F and WF phases \cite{chen2014nonlinear}, resulting in a flat dispersion (see the SM for an example). This phenomenon can be compared to the $\phi^4$ model as it approaches its so-called anti-continuum limit \cite{chirilus2014sine} ($C\rightarrow0$ in our case; see the SM for the dispersion relation of an infinite homogeneous $\phi^4$ chain).
We also observe internal modes for the kink states of both chains. 
As expected, the KL chain exhibits a zero mode (to $\tilde{\omega}^2<10^{-7}$) that does not change with kink width for both intersite- and onsite-centered kink states, as predicted by the Maxwell–Calladine index theorem \cite{calladine1978buckminster,sun2012surface,kane2014topological,mao2018maxwell}.
In addition to the zero mode in the KL chain, there are extra finite-frequency modes present in the band gap of the spectrum of the kink state for both chains. These modes are spatially localized around the kink, and are the aforementioned internal modes \cite{kevrekidis2000dynamics,dauxois2006physics,chirilus2014sine,kevrekidis2019dynamical}. 
For the discrete $\phi^4$ chain, in Fig.~\ref{fig:kink_dynamics}(i), we can see that the lowest frequency internal mode (LFIM) decreases nearly linearly on a logarithmic scale as the kink width increases for $\tilde{w}_0>1$. This approach toward vanishingly low energy is consistent with the expected zero mode present in the continuum $\phi^4$ kink \cite{kevrekidis2000dynamics,dauxois2006physics,chirilus2014sine,kevrekidis2019dynamical}. 
Furthermore, we find that as the F-phase kink approaches its continuum limit, only a single internal mode persists. Its frequency asymptotically converges to $\tilde{\omega}_s^2=3/4$, which corresponds to the shape mode (a type of internal mode) of the continuum $\phi^4$ kink \cite{kevrekidis2000dynamics,dauxois2006physics,chirilus2014sine,kevrekidis2019dynamical}. This further supports the mapping between the continuum F-phase kink and the continuum $\phi^4$ field theory, as introduced in Ref.~\cite{chen2014nonlinear}.  
Besides that, we observe that as $\tilde{w}_0\rightarrow 0$, extra internal modes emerge within the band gap of the KL chain's kink, whereas all modes of the intersite-centered $\phi^4$ kink collapse into a single frequency in the aforementioned anti-continuum limit.
See the SM for zoomed-in comparison of internal mode frequencies in F-phase kinks and $\phi^4$ kinks for a broader kink width range, as well as comparison of the mode shapes of the non-zero internal modes for selected kinks with $\tilde{w}_0<1.6$.

Next, we turn our attention to the energy barrier that the kinks (both in KL chain and discrete $\phi^4$ chain) must overcome in order to initiate motion. To characterize the energy barriers, we examine the short-time dynamics of kinks by assigning initial velocities to each site in the kink state, proportional to the eigenmode amplitudes of the zero mode in the KL chain and the LFIM in the discrete $\phi^4$ chain (as shown in Fig.~\ref{fig:kink_dynamics}(c,f) and (i), respectively). Examples of these initial velocities are shown in the insets of Fig.~\ref{fig:kink_dynamics}(d,g,j). Using these velocities, we characterize the input energy as $\sum_n{mr^2\dot{\psi}_{n,t=0}^2/2}$ for the KL chain and $\sum_n{m\dot{u}_{n,t=0}^2/2}$ for the discrete $\phi^4$ chain.
The kink velocity is determined by measuring the time it takes for the kink's center to shift by one site---specifically, when the $141$st rotor (or particle) achieves the same angle (or displacement) as the $140$th rotor (or particle).
Figure~\ref{fig:kink_dynamics}(d,g) shows no energy barrier for the KL chain kink, as evidenced by the linear relationship between input energy and kink velocity, and the absence of a cutoff point. This confirms the absence of the PN barrier for the kink in the KL chain.
In contrast, in Fig.~\ref{fig:kink_dynamics}(j), as the $\phi^4$ kink becomes more discrete and the kink narrows (with narrower kinks denoted by darker blue shading), cutoff points appear at greater and greater input energies, below which the input energy is insufficient to achieve a kink velocity above the resolution limit.


We now shift our focus to another key feature of discrete kink-supporting systems.
As noted above, a general property of these systems, including discrete $\phi^4$ chains, is that they support only two stationary kink solutions: a stable intersite-centered kink and an unstable onsite-centered kink \cite{kivshar1993peierls,braun2004frenkel,chirilus2014sine,kevrekidis2019dynamical}.
In contrast, we find that the KL chain supports an infinite set of stationary kink solutions due to the presence of a zero mode. Here, the kink solutions between two sites are obtained through a low-speed dynamical simulation, with the initial on-site configuration determined using the Newton-Raphson method (see the SM for more details). In Fig.~\ref{fig:kink_dynamics}(e), we show an example featuring the same kink state that we investigate in our experiments ($\tilde{r}=1.5$, $d_{\mathrm{exp}}=2.0296$), where, as the kink's center shifts between two sites, the zero mode consistently exists and the internal-modes undergo a continuous and dramatic variation in their frequencies. 
Interestingly, we further find that the smoothness of this internal-mode frequency variation depends on $d$ for fixed $\tilde{r}$, where larger $d$ leads to sharper changes in internal-mode frequencies within the band gap (see the SM for examples).

\subsection{Simulated phonon-kink interaction}

\begin{figure*}[ht!]
    \centering
    \includegraphics[width=17.2cm]{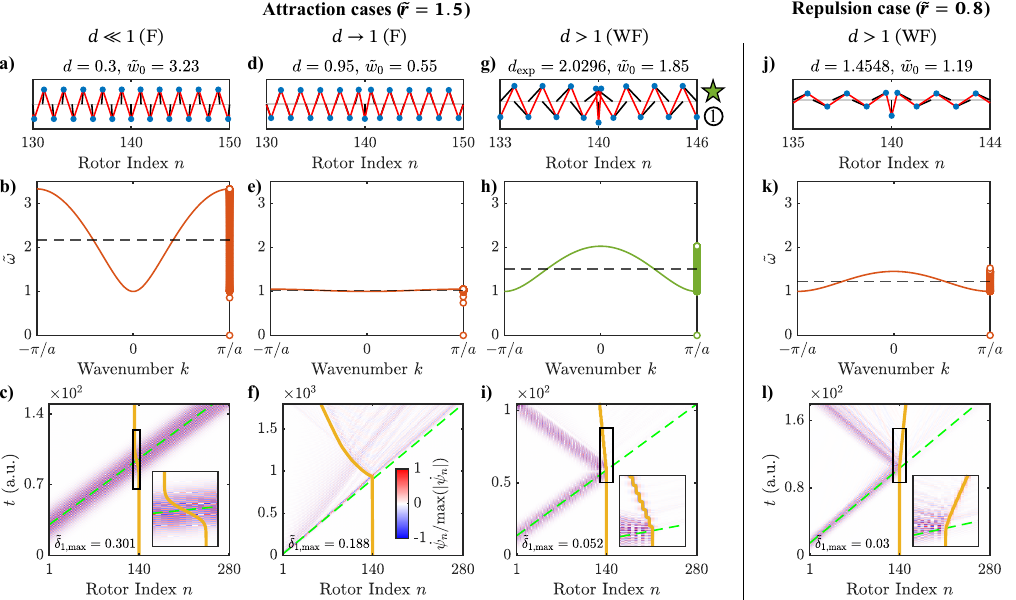}
    \caption{
    \textbf{Simulated phonon-kink interactions in KL chains with different unit cell geometries.}
    (Top row) Zoomed-in view of the onsite-centered kink state ($\psi_{140}=0$) for KL chains with $280$ rotors.
    (Middle row) Dispersion relation of homogeneous chains with finite eigenmodes overlapped at $k=\pi/a$ for each case in the top row. The black dashed line denotes the center frequency of the phonon wave packet, which is in the middle of the phonon band.
    (Bottom row) Spatiotemporal angular velocity response, with green dashed lines denoting the predicted center position of the phonon wave packet based on the group velocity in the infinite homogeneous chain and yellow solid lines indicating the fitted center position of the kink. The first three columns from the left denote $\tilde{r} = 1.5$ at different $d$ in the F/WF phase, and the rightmost $\tilde{r} = 0.8$ in the WF phase. \textcircled{1} indicates the experimental onsite-centered kink state shown in Fig.~\ref{fig:kink_dynamics}(e).}
    \label{fig:longtime_phonon-kink_dynamics_1}
\end{figure*}

\begin{figure*}[ht!]
    \centering
    \includegraphics[width=17.2cm]{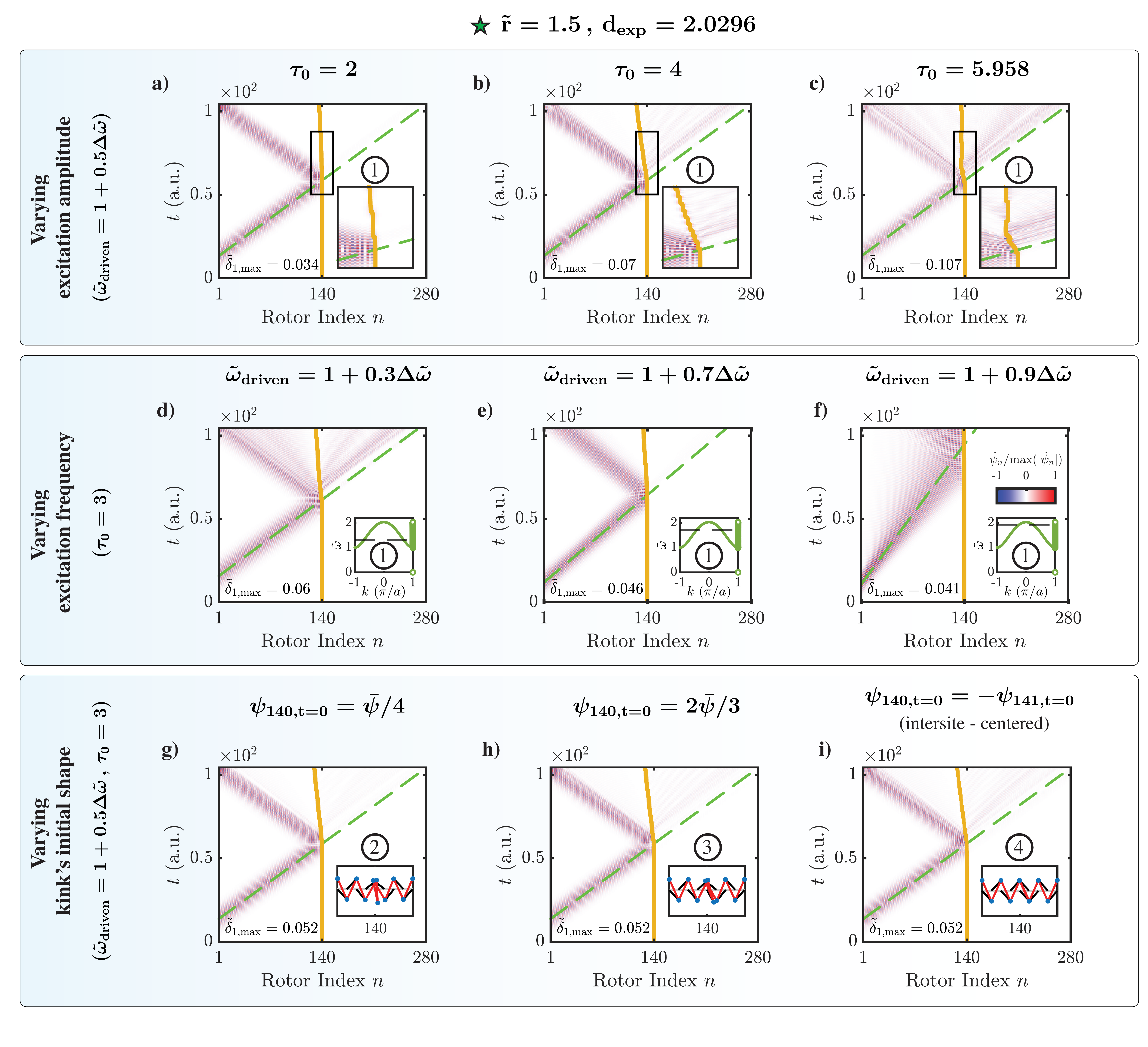}
    \caption{
    \textbf{Simulated phonon-kink interaction in KL chains with the experimental unit cell geometry.}
    (Top row) Comparison of kink dynamics under varying excitation amplitudes at a constant mid-band excitation frequency. (Middle row) Comparison of kink dynamics under different excitation frequencies (indicated by black dashed lines in the insets) at fixed excitation torque amplitude.
    (Bottom row) Comparison of kink dynamics at different kink's center positions, under a constant mid-band excitation frequency and excitation torque amplitude.
    Circled numbers indicate the kink states shown in Fig.~\ref{fig:kink_dynamics}(e). $\Delta\tilde{\omega}$ represents the normalized phonon band width, defined as $\abs{\tilde{\omega}_{k=0}-\tilde{\omega}_{k=\pi/a}}$.
    Green dashed lines denote the predicted center position of the phonon wave packet based on the group velocity in the infinite homogeneous chain and yellow solid lines indicate the fitted center position of the kink.}\label{fig:longtime_phonon-kink_dynamics_2}
\end{figure*}

Having calculated the dynamical features of kinks in the KL chain, in this section, we computationally explore the dynamics of phonon interactions with F/WF-phase kinks. 
In Fig.~\ref{fig:longtime_phonon-kink_dynamics_1}, we show the dependence of the phonon-kink interaction dynamics on the unit cell geometric parameters $d$ and $\tilde{r}$. In Fig.~\ref{fig:longtime_phonon-kink_dynamics_2}, we show the dependence of phonon-kink interaction dynamics on wave packet amplitude, wave packet center frequency, and the initial kink's center position, for chains with the same unit cell geometric parameters as used in our experiments. 

Considering first Fig.~\ref{fig:longtime_phonon-kink_dynamics_1}, the top row shows a zoomed-in view of the kink (for the specified, varied, unit cell geometric parameters), the middle row the corresponding phonon band, and the bottom row the spatiotemporal chain response during the phonon wave packet generation, kink interaction, and subsequent scattering. 
For these phonon-kink interaction simulations, we use the Newton-Raphson method to solve for the onsite-centered kink state comprising the chain's initial configuration, then inject the wave packet by applying a Gaussian-modulated sinusoidal torque $\tau(t)=\tau_0 e^{-(t-t_0)^2/(2\sigma_t^2)}\sin{\omega_{\mathrm{driven}} t}$ to the left edge (the first rotor), with $t_0 = 40\pi/\omega_{\mathrm{driven}}$, $\sigma_t =12\pi/\omega_{\mathrm{driven}}$, $\omega_{\mathrm{driven}}$ the excitation frequency centered in the middle of the phonon band, and $\tau_0$ set to $0.5769$, $0.17$, $3$, and $0.5241$ for the columns from left to right in Fig.~\ref{fig:longtime_phonon-kink_dynamics_1}, respectively. 
We denote the normalized amplitude of the phonon wave packet for each case in the bottom row. This amplitude is defined as $\tilde{\delta}_{1,\mathrm{max}}\equiv\max{\left(\abs{\left(\sin{\psi_1(t\leq2t_0)}-\sin{\psi_{1,t=0}}\right)/\sin{\bar{\psi}}}\right)}$ (see Ref.~\cite{chen2014nonlinear} for the rationale behind the choice of the sine function), which represents the first rotor's maximum $x$-deflection relative to its mass position in the homogeneous chain within a $2t_0$ time range. 

We first describe three cases corresponding to $\tilde{r} = 1.5$, with $d = 0.3$, $0.95$, and $2.0296$ (leftmost to third-from-the-left columns in Fig.~\ref{fig:longtime_phonon-kink_dynamics_1}, respectively), wherein the phonon-kink interaction is attractive \textit{i.e.} the kink's velocity is opposite to the velocity of the incoming phonon wave packet. We observe that, when $d$ is small (leftmost column), the kink only moves during the interaction with the phonons (Fig.~\ref{fig:longtime_phonon-kink_dynamics_1}(c)) and there is minimal scattering of phonons. This behavior is similar to that of the $\phi^4$ kinks close to their continuum limit \cite{hasenfratz1977interaction}, which is consistent, as small $d$ approaches the continuum limit that resembles the $\phi^4$ field theory \cite{chen2014nonlinear}.
As $d\rightarrow1$ (second column from the left in Fig.~\ref{fig:longtime_phonon-kink_dynamics_1}), the kink becomes highly discrete ($\tilde{w}_0\rightarrow0$), the rotors and springs align more closely, and the phonon band narrows, which leads to strong dispersion of the phonons before they reach the kink. As a result, in Fig.~\ref{fig:longtime_phonon-kink_dynamics_1}(f), although initially the kink moves rapidly in the direction from whence phonons came, it soon slows during the time it is interacting with the later-arriving, dispersed phonons. We also observe phonons propagating away from the kink as it moves, which may constitute a combination of phonon reflection, scattering, and radiation from the kink.
For $d>1$ (third column from the left in Fig.~\ref{fig:longtime_phonon-kink_dynamics_1}), the kink transitions into a WF-phase kink, where the kink width increases with $d$, but remains relatively small, and the phonon band widens, leading to reduced phonon dispersion.
In this case, after the phonons hit the kink, most are reflected (unlike the $d\rightarrow0$, continuum limit, case), with a small amount of phonons seemingly radiated from the kink in both directions. We also see in the inset of Fig.~\ref{fig:longtime_phonon-kink_dynamics_1}(i) that the kink has a somewhat oscillatory trajectory, which can also be observed in the discrete $\phi^4$ chain (see the SM).
The kink also continues to move even after the phonons move away from it, similar to the discrete $\phi^4$ kink behavior \cite{theodorakopoulos1980lattice}. However, the resulting kink dynamics are not identical between the two systems, as the presence of the PN barrier in the discrete $\phi^4$ chain causes the kink to eventually slow down and stop due to energy radiating from the kink as it moves (see the SM for a simulated example), an effect which is not present for the KL chain kinks. Simulation of phonon-kink interaction in the $\phi^4$ chain close to its continuum limit is also provided in the SM for comparison.

By tuning the geometrical parameters ($\tilde{r}$), we observe that the KL chain also supports repulsive interactions between the kink and the phonons. In the rightmost column of Fig.~\ref{fig:longtime_phonon-kink_dynamics_1}, we show a repulsion case, corresponding to $\tilde{r} = 0.8$ and $d = 1.4548$, which is also a WF-phase kink. Similar to the attraction case in the third column, we observe that most phonons are reflected after hitting the kink, and the kink continues to move even after the phonons have moved away from it. In this case, the kink moves in the opposite direction, exhibiting repulsive behavior. While kink attraction via phonon interaction is the more common phenomenon \cite{hasenfratz1977interaction,theodorakopoulos1980lattice}, repulsive behavior during phonon-kink interaction has also been observed computationally in discrete $\phi^4$ \cite{abdelhady2011wave} and $\phi^6$ \cite{saadatmand2024phonons} chains. In addition, we observe stronger phonon radiation as the kink moves, particularly shortly after the interaction. In the SM, we provide a first-order 
perturbation theory calculation that predicts the type of phonon-kink interaction shortly after the phonon reaches the kink.

Noting that the results shown in Fig.~\ref{fig:longtime_phonon-kink_dynamics_1} are not directly comparable due to the different excitation frequencies and amplitudes involved, as well as the fact that the interaction is a nonlinear process, we next choose the kink state that we will study experimentally and examine (Fig.~\ref{fig:longtime_phonon-kink_dynamics_2}) the influence of wave packet amplitude (top row), wave packet center frequency (middle row), as well as the initial kink's center position (bottom row). 
The excitation is defined in the same way as in Fig.~\ref{fig:longtime_phonon-kink_dynamics_1}. 
In the top row, we observe that as the driving torque increases ($\tau_0 = 2$, $4$, and $5.958$), the kink moves faster with more phonons radiated opposite to the kink's direction of motion (Fig.~\ref{fig:longtime_phonon-kink_dynamics_2}(a,b), wherein the kinks have velocities of $-0.0885$ and $-0.3118$, respectively), until eventually the highest amplitude shown results in chaotic kink movements accompanied by significant phonon radiation from the kink (Fig.~\ref{fig:longtime_phonon-kink_dynamics_2}(c)). 
As concerns the dependence of the interaction on frequency of the phonon wave packet within the propagating band (middle row), with fixed torque amplitude ($\tau_0=3$), we observed: significantly differing amounts of dispersal of the phonon wave packet (as expected due to the differing local curvature of the dispersion curve and proximity to the band edges), differing amounts of phonon radiation from the kink, and minimally modified (compared to kink velocity changes due to phonon wave packet amplitude) kink velocity. For the three cases tested, the kink velocity ranged from $-0.1695$ (Fig.~\ref{fig:longtime_phonon-kink_dynamics_2}(d)) to $-0.0418$ (Fig.~\ref{fig:longtime_phonon-kink_dynamics_2}(f)).  
Finally, as concerns effects of the kink's center position on the phonon-kink interaction dynamics (bottom row, $\tau_0=3$), we observe different amounts of phonon radiation from the kink (following the wave packet interaction) and minimal changes in kink velocity ($-0.2366$ to $-0.2864$). 
We suggest this breadth of nonlinear dynamical phenomena demonstrated in our simulation of phonon-kink interaction in the KL chain motivates future studies delving into the mechanisms underlying kink response as a function of phonon amplitude, phonon frequency, and kink's center position, as well as their interdependency with the underlying linear dynamical features including the phonon spectrum of the homogeneous chain and the internal modes of the zero-energy kink. 

\subsection{Experimental observation of kink control and generation via phonons}

\begin{figure*}[ht!]
    \centering
    \includegraphics[width=17.2cm]{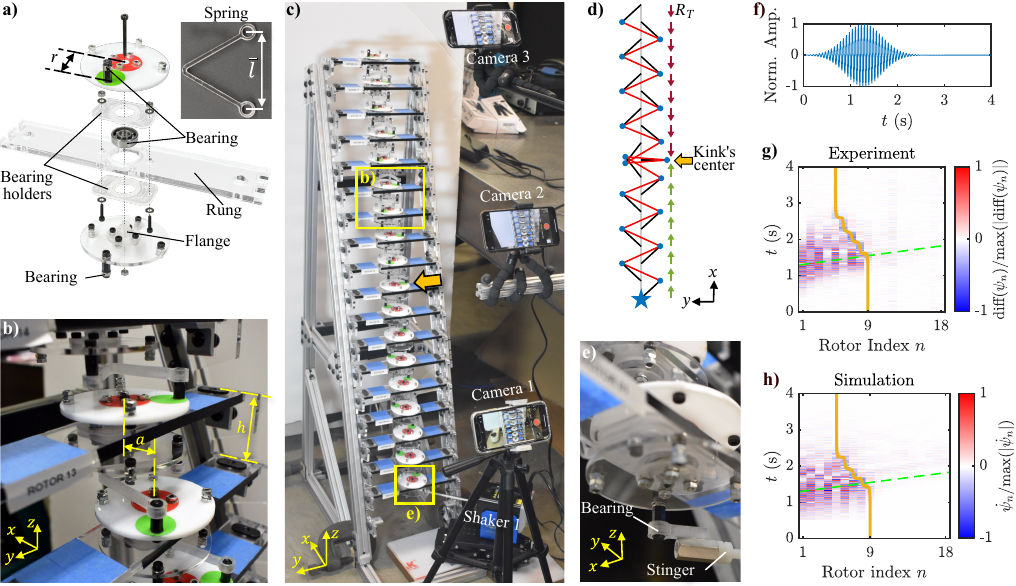}
    \caption{
    \textbf{Experimental observation of phonon-kink interaction (kink control) in the KL chain.}
    (a) Single rotor assembly. Inset shows a single polycarbonate spring.
    (b) Two rotors connected via polycarbonate springs.
    (c) Experimental setup with the chain of $18$ rotors configured as in (d).
    (d) Chain configuration with the center of the kink localized at the $9$th rotor (yellow arrow) with green and red arrows indicating the polarization vectors. The blue star marks the excited rotor ($n=1$).
    (e) Bottom view of the excited rotor.
    (f) Normalized excitation function.
    Experimentally measured (g) and simulated (h) spatiotemporal angular velocity response with green dashed lines denoting the predicted wave packet center position based on the homogeneous group velocity and yellow solid lines indicating the kink's fitted center position. 
    }\label{fig:kink_control}
\end{figure*}

To experimentally realize the described KL chain supporting elastic waves, 
we construct a rotor chain, in which each rotor is coupled by bent thin beams as effective springs. We choose parameters $a = 20$ mm, $r = 30$ mm, and $\bar{l} = 48.5$ mm, which yield $\Bar{\psi}=0.7431$ rad and $d_{\mathrm{exp}} = 2.0296$, and place the zero-energy kink in this chain within the WF phase with a small kink width ($\tilde{w}_0=1.85$). We avoid selecting discrete kinks in the F phase because the configuration space is narrow, which makes the kink's properties highly sensitive to geometric accuracy, as suggested in the phase diagram in Fig.~\ref{fig:kink_dynamics}(b).
We machine the springs from polycarbonate sheets, with the parameters and design detailed in the SM. The springs feature small circular pockets at the ends for ball bearings, which minimize moments during rotation. The effective normal stiffness of the spring is measured to be $384.3$ N/m. 
Inspired by the LEGO-based systems demonstrated in Refs.~\cite{chen2014nonlinear,upadhyaya2020nuts}, we design the single rotor as a two-sided reel, as shown in Fig.~\ref{fig:kink_control}(a). Red and green circular markers on the top disc indicate the location of the disc's center and the spring's bearing, respectively. Mass is evenly distributed on all the discs to avoid wobble and the moment of inertia of an entire rotor is $2.97\times10^{-5}$ kg$\cdot$m$^2$ (see the SM for more details of the rotors). The central ball bearing of the rotor is installed on a horizontal rung. By staggering rungs with a relative out-of-plane (${z}$) distance ($h = 55$ mm, Fig.~\ref{fig:kink_control}(b)), overlapping configurations of the KL chain are enabled. 
Figure~\ref{fig:kink_control}(c) shows the experimental KL chain consisting of $18$ rotors. 

\subsubsection{Kink control with phonons}

To initialize the configuration with a kink in the middle of the chain, we first set the rotor angles to those of the homogeneous state, then form and move the kink from the soft edge by applying large manual deformations to its center region until the kink reaches the desired central position (Fig.~\ref{fig:kink_control}(c)), where $\psi_{9}=0$ (Fig.~\ref{fig:kink_control}(d)). 
To inject phonons into the chain, an electrodynamic shaker (shaker 1 in Fig.~\ref{fig:kink_control}(c)) is coupled to the bottom edge of the chain (the first rotor) via a ball bearing and a nylon stinger, which provides a tangential linear excitation (Fig.~\ref{fig:kink_control}(e)). 
The excitation voltage signal to the shaker (via an amplifier) follows the same form as the Gaussian-modulated sinusoidal torque used in simulations, with $\omega_{\mathrm{driven}}=2\pi f$ and $f = 15.65$ Hz (the median of the eigenfrequencies calculated from the chain configuration shown in Fig.~\ref{fig:kink_control}(d)). The normalized excitation signal is shown in Fig.~\ref{fig:kink_control}(f). We track the dynamic response of the chain using digital image processing of synchronized videos recorded at $240$ fps by three iPhones, approximately equidistantly spaced along the chain, as shown in Fig.~\ref{fig:kink_control}(c) (see the SM for details of data acquisition). 

In Fig.~\ref{fig:kink_control}(g), the kink can be seen to move upon interaction with the phonon wave packet. However, in contrast to the expected, simulated continuous kink motion following phonon-kink interaction (see the SM also for simulated demonstration of this continuous motion in a longer chain with the same unit cell parameters as in the experiment), we observe that the kink stops moving after shifting about four sites. We believe this behavior can be attributed to damping. In Fig.~\ref{fig:kink_control}(g,h), we compare our experimental result with numerical simulation, as before (\textit{e.g.}, bottom row of Fig.~\ref{fig:longtime_phonon-kink_dynamics_1}), but now with the inclusion of an experimentally determined onsite viscous damping (see the SM for details of damping characterization). 
We observe that the kink moves toward the source of the excitation after initial phonon-kink interaction, with excellent agreement between the experimentally measured and numerically predicted kink trajectory, as well as phonon reflection and radiation. The  observed phenomena also agree with the attractive behavior that we predicted in the third column of Fig.~\ref{fig:longtime_phonon-kink_dynamics_1} and Fig.~\ref{fig:longtime_phonon-kink_dynamics_2} for $\tilde{r} = 1.5$ and $d_{\mathrm{exp}} = 2.0296$. To verify that the motion of the kink is not due to gravity (since it moves downward), we repeated the test by exciting the top edge of the chain ($18$th rotor) and observed that phonons indeed attract the kink (see the SM Video 1 and 2 for exciting bottom and top edge, respectively). 
We also excited both ends of the chain at a frequency in the band gap ($5$ Hz) and did not observe any movement of the kink as expected (see the SM Video 3 and 4 for exciting bottom and top edge, respectively).

\begin{figure}[h]
    \centering
    \includegraphics[width=8.6cm]{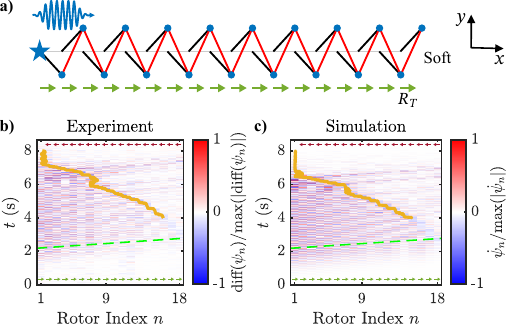}
    \caption{
    \textbf{Experimental observation of kink generation via phonons in the KL chain.}
    (a) Homogeneous chain configuration with green arrows indicating the initial polarization vectors (RPS). The blue star marks the excited rotor ($n = 1$). 
    (b,c) Experimentally measured (b) and simulated (c) spatiotemporal angular velocity response with green dashed lines denoting the predicted ramped peak of the excitation based on the homogeneous group velocity and yellow solid lines indicating the kink's fitted center position from $4$ to $8$ s. Small arrows in (b,c) indicate the initial (green) and final (red) polarization vectors.}
    \label{fig:kink_generation}
\end{figure}

\subsubsection{Kink generation with phonons}

We observe that a kink can be generated from the soft edge of an initially homogeneous chain by injecting phonons from the opposite, rigid edge. Although it has been reported that kink generation and propagation can be initiated in active systems \cite{woodhouse2018autonomous, ghatak2020observation, veenstra2024non}, we demonstrate it can be realized via only passive elements. 
By actuating the rigid edge of the same experimental chain in the homogeneous state (Fig. \ref{fig:kink_generation}(a)) with a longer, ramped sinusoidal excitation with a frequency of $12.7$ Hz, which resides in the phonon band, we observe the formation and propagation of the kink as it moves and accelerates across the chain, switching the polarization of the bulk, as shown in Fig.~\ref{fig:kink_generation}(b) and (c). We note that the kink is generated by phonons and remains mobile, due to the longer wave packet duration, even in the presence of dissipation. 
Again, the experimentally measured (Fig.~\ref{fig:kink_generation}(b)) and simulated (Fig.~\ref{fig:kink_generation}(c), including damping) kink trajectory and phonon scattering behavior show excellent agreement (see the SM for details on the excitation signal setting and spatiotemporal strain response for both experiment and simulation). Again, we observe that most of the phonons are reflected by the kink. We also provide additional tests with excitation outside the phonon band ($6.35$ Hz and $25.4$ Hz), where no kink movement was observed for these cases (see the SM Videos 5-7 for these tests).
In the SM, we further simulate a repulsion case of the kink in the WF phase for parameters corresponding to the experiment, both with and without onsite damping.  
While the repulsion can be observed for the case without damping for these parameters, we were unable to observe it for the case with damping. Increasing the driving amplitude for the damped case in an attempt to observe such repulsion, resulted in the formation of a chaotic dynamical state before repulsion was seen. Because of this, we did not attempt to experimentally observe repulsion.

\section{Conclusion}

In this work, we reported the first experimental observation of mechanical kink control and generation via phonons, independently of the kink's width (or equivalently its discreteness).
In addition to the experimental realization in our newly introduced, elastically coupled KL chain, we numerically observed a continuously varying family of kink solutions in contrast to the onsite- and intersite-centered kinks of other nonlinear discrete systems. Simulations showed distinct phonon-kink dynamics unique to the KL chain, including both attraction and repulsion, as well as long-duration, post-interaction kink motion. 

Building upon this new capacity for kink control in lattices supporting zero modes, we anticipate many application-relevant benefits including robust, topologically protected signal transmission (in the form of kinks, and with application to signal processing and logic) and remote control of localized, anisotropic, and extreme material stiffness. Other future potential application areas include locomotion and shape-shifting materials, as well as sensing (\textit{e.g.}, large signal kink control, with small, phonon-like stimuli). 

Future related research directions include extending our system into higher dimensions and smaller scales, as well as answering fundamental questions regarding kink dynamics in the KL chain. Regarding higher dimensions, several works have studied the existence of domain walls in 2D topological metamaterials that support zero-energy modes \cite{paulose2015topological,rocklin2017transformable}. Regarding scalability, we anticipate possibilities in scale reduction, consider prior demonstration of rotor-like behaviors in systems such as bacterial flagellar motors \cite{nirody2017biophysicist,singh2024cryoem}, DNA \cite{shi2022sustained}, and nanoelectromechanics \cite{kim2014ultrahigh}. Given the importance of highly-discrete kinks in the form of dislocations in condensed matter \cite{dauxois2006physics}, one might ask ``what are the thermal properties of materials supporting highly discrete kinks that lack a PN barrier?'' Several near-term, future questions regarding kink dynamics in the KL chain are also stimulated by our results. What are the dynamics of the antikink in the KL chain \cite{zhou2017kink,upadhyaya2020nuts}, noting that they are not zero-energy and have a finite PN barrier? This includes their interaction with moving zero-energy kinks as well as antikink-phonon interaction. What are kink dynamics in a quasi-periodic KL chain \cite{sato2018solitons}? Finally, is there a relationship between the internal mode evolution with the kink's center position between sites and the short-time attraction and repulsive behavior? Given these applied and fundamental possibilities, we anticipate fruitful extensions of discreteness-independent, barrier-free phonon-kink interaction dynamics. \\

\begin{acknowledgments}

K.Q., N.C., X.M., and N.B. acknowledge support from the US Army Research Office (Grant No. W911NF-20-2-0182). 
N.C., F.S., K.S., and X.M. acknowledge support from the US Office of Naval Research (MURI N00014-20-1-2479). 
This research is funded in part by a grant from ICAM the Institute for Complex Adaptive Matter to K.Q. 
Additionally, K.Q. acknowledges support from the UCSD MAE Stanford S. `Sol' Penner Post-Doctoral Research Travel Award.

\end{acknowledgments}


\begin{thebibliography}{96}%
\makeatletter
\providecommand \@ifxundefined [1]{%
 \@ifx{#1\undefined}
}%
\providecommand \@ifnum [1]{%
 \ifnum #1\expandafter \@firstoftwo
 \else \expandafter \@secondoftwo
 \fi
}%
\providecommand \@ifx [1]{%
 \ifx #1\expandafter \@firstoftwo
 \else \expandafter \@secondoftwo
 \fi
}%
\providecommand \natexlab [1]{#1}%
\providecommand \enquote  [1]{``#1''}%
\providecommand \bibnamefont  [1]{#1}%
\providecommand \bibfnamefont [1]{#1}%
\providecommand \citenamefont [1]{#1}%
\providecommand \href@noop [0]{\@secondoftwo}%
\providecommand \href [0]{\begingroup \@sanitize@url \@href}%
\providecommand \@href[1]{\@@startlink{#1}\@@href}%
\providecommand \@@href[1]{\endgroup#1\@@endlink}%
\providecommand \@sanitize@url [0]{\catcode `\\12\catcode `\$12\catcode `\&12\catcode `\#12\catcode `\^12\catcode `\_12\catcode `\%12\relax}%
\providecommand \@@startlink[1]{}%
\providecommand \@@endlink[0]{}%
\providecommand \url  [0]{\begingroup\@sanitize@url \@url }%
\providecommand \@url [1]{\endgroup\@href {#1}{\urlprefix }}%
\providecommand \urlprefix  [0]{URL }%
\providecommand \Eprint [0]{\href }%
\providecommand \doibase [0]{https://doi.org/}%
\providecommand \selectlanguage [0]{\@gobble}%
\providecommand \bibinfo  [0]{\@secondoftwo}%
\providecommand \bibfield  [0]{\@secondoftwo}%
\providecommand \translation [1]{[#1]}%
\providecommand \BibitemOpen [0]{}%
\providecommand \bibitemStop [0]{}%
\providecommand \bibitemNoStop [0]{.\EOS\space}%
\providecommand \EOS [0]{\spacefactor3000\relax}%
\providecommand \BibitemShut  [1]{\csname bibitem#1\endcsname}%
\let\auto@bib@innerbib\@empty
\bibitem [{\citenamefont {Dauxois}\ and\ \citenamefont {Peyrard}(2006)}]{dauxois2006physics}%
  \BibitemOpen
  \bibfield  {author} {\bibinfo {author} {\bibfnamefont {T.}~\bibnamefont {Dauxois}}\ and\ \bibinfo {author} {\bibfnamefont {M.}~\bibnamefont {Peyrard}},\ }\href@noop {} {\emph {\bibinfo {title} {Physics of solitons}}}\ (\bibinfo  {publisher} {Cambridge University Press},\ \bibinfo {year} {2006})\BibitemShut {NoStop}%
\bibitem [{\citenamefont {Vachaspati}(2007)}]{vachaspati2007kinks}%
  \BibitemOpen
  \bibfield  {author} {\bibinfo {author} {\bibfnamefont {T.}~\bibnamefont {Vachaspati}},\ }\href@noop {} {\emph {\bibinfo {title} {Kinks and domain walls: An introduction to classical and quantum solitons}}}\ (\bibinfo  {publisher} {Cambridge University Press},\ \bibinfo {year} {2007})\BibitemShut {NoStop}%
\bibitem [{\citenamefont {Chaikin}\ \emph {et~al.}(1995)\citenamefont {Chaikin}, \citenamefont {Lubensky},\ and\ \citenamefont {Witten}}]{chaikin1995principles}%
  \BibitemOpen
  \bibfield  {author} {\bibinfo {author} {\bibfnamefont {P.~M.}\ \bibnamefont {Chaikin}}, \bibinfo {author} {\bibfnamefont {T.~C.}\ \bibnamefont {Lubensky}},\ and\ \bibinfo {author} {\bibfnamefont {T.~A.}\ \bibnamefont {Witten}},\ }\href@noop {} {\emph {\bibinfo {title} {Principles of condensed matter physics}}},\ Vol.~\bibinfo {volume} {10}\ (\bibinfo  {publisher} {Cambridge university press Cambridge},\ \bibinfo {year} {1995})\BibitemShut {NoStop}%
\bibitem [{\citenamefont {Frenkel}(1939)}]{frenkel1939theory}%
  \BibitemOpen
  \bibfield  {author} {\bibinfo {author} {\bibfnamefont {J.}~\bibnamefont {Frenkel}},\ }\bibfield  {title} {\bibinfo {title} {On the theory of plastic deformation and twinning},\ }\href@noop {} {\bibfield  {journal} {\bibinfo  {journal} {J. Phys.}\ }\textbf {\bibinfo {volume} {1}},\ \bibinfo {pages} {137} (\bibinfo {year} {1939})}\BibitemShut {NoStop}%
\bibitem [{\citenamefont {Weiner}(1964)}]{weiner1964dislocation}%
  \BibitemOpen
  \bibfield  {author} {\bibinfo {author} {\bibfnamefont {J.}~\bibnamefont {Weiner}},\ }\bibfield  {title} {\bibinfo {title} {Dislocation velocities in a linear chain},\ }\href@noop {} {\bibfield  {journal} {\bibinfo  {journal} {Physical Review}\ }\textbf {\bibinfo {volume} {136}},\ \bibinfo {pages} {A863} (\bibinfo {year} {1964})}\BibitemShut {NoStop}%
\bibitem [{\citenamefont {Atkinson}\ and\ \citenamefont {Cabrera}(1965)}]{atkinson1965motion}%
  \BibitemOpen
  \bibfield  {author} {\bibinfo {author} {\bibfnamefont {W.}~\bibnamefont {Atkinson}}\ and\ \bibinfo {author} {\bibfnamefont {N.}~\bibnamefont {Cabrera}},\ }\bibfield  {title} {\bibinfo {title} {Motion of a frenkel-kontorowa dislocation in a one-dimensional crystal},\ }\href@noop {} {\bibfield  {journal} {\bibinfo  {journal} {Physical Review}\ }\textbf {\bibinfo {volume} {138}},\ \bibinfo {pages} {A763} (\bibinfo {year} {1965})}\BibitemShut {NoStop}%
\bibitem [{\citenamefont {Earmme}\ and\ \citenamefont {Weiner}(1974)}]{earmme1974breakdown}%
  \BibitemOpen
  \bibfield  {author} {\bibinfo {author} {\bibfnamefont {Y.~Y.}\ \bibnamefont {Earmme}}\ and\ \bibinfo {author} {\bibfnamefont {J.}~\bibnamefont {Weiner}},\ }\bibfield  {title} {\bibinfo {title} {Breakdown phenomena in high-speed dislocations},\ }\href@noop {} {\bibfield  {journal} {\bibinfo  {journal} {Journal of Applied Physics}\ }\textbf {\bibinfo {volume} {45}},\ \bibinfo {pages} {603} (\bibinfo {year} {1974})}\BibitemShut {NoStop}%
\bibitem [{\citenamefont {Flytzanis}\ \emph {et~al.}(1977)\citenamefont {Flytzanis}, \citenamefont {Crowley},\ and\ \citenamefont {Celli}}]{flytzanis1977solitonlike}%
  \BibitemOpen
  \bibfield  {author} {\bibinfo {author} {\bibfnamefont {N.}~\bibnamefont {Flytzanis}}, \bibinfo {author} {\bibfnamefont {S.}~\bibnamefont {Crowley}},\ and\ \bibinfo {author} {\bibfnamefont {V.}~\bibnamefont {Celli}},\ }\bibfield  {title} {\bibinfo {title} {Solitonlike motion of a dislocation in a lattice},\ }\href@noop {} {\bibfield  {journal} {\bibinfo  {journal} {Physical Review Letters}\ }\textbf {\bibinfo {volume} {39}},\ \bibinfo {pages} {891} (\bibinfo {year} {1977})}\BibitemShut {NoStop}%
\bibitem [{\citenamefont {Komarova}\ and\ \citenamefont {Soffera}(2005)}]{komarova2005nonlinear}%
  \BibitemOpen
  \bibfield  {author} {\bibinfo {author} {\bibfnamefont {N.~L.}\ \bibnamefont {Komarova}}\ and\ \bibinfo {author} {\bibfnamefont {A.}~\bibnamefont {Soffera}},\ }\bibfield  {title} {\bibinfo {title} {Nonlinear waves in double-stranded dna},\ }\href@noop {} {\bibfield  {journal} {\bibinfo  {journal} {Bulletin of mathematical biology}\ }\textbf {\bibinfo {volume} {67}},\ \bibinfo {pages} {701} (\bibinfo {year} {2005})}\BibitemShut {NoStop}%
\bibitem [{\citenamefont {Su}\ \emph {et~al.}(1979)\citenamefont {Su}, \citenamefont {Schrieffer},\ and\ \citenamefont {Heeger}}]{su1979solitons}%
  \BibitemOpen
  \bibfield  {author} {\bibinfo {author} {\bibfnamefont {W.-P.}\ \bibnamefont {Su}}, \bibinfo {author} {\bibfnamefont {J.~R.}\ \bibnamefont {Schrieffer}},\ and\ \bibinfo {author} {\bibfnamefont {A.~J.}\ \bibnamefont {Heeger}},\ }\bibfield  {title} {\bibinfo {title} {Solitons in polyacetylene},\ }\href@noop {} {\bibfield  {journal} {\bibinfo  {journal} {Physical review letters}\ }\textbf {\bibinfo {volume} {42}},\ \bibinfo {pages} {1698} (\bibinfo {year} {1979})}\BibitemShut {NoStop}%
\bibitem [{\citenamefont {Bruce}(1981)}]{bruce1981scattering}%
  \BibitemOpen
  \bibfield  {author} {\bibinfo {author} {\bibfnamefont {D.}~\bibnamefont {Bruce}},\ }\bibfield  {title} {\bibinfo {title} {Scattering properties of ferroelectric domain walls},\ }\href@noop {} {\bibfield  {journal} {\bibinfo  {journal} {Journal of Physics C: Solid State Physics}\ }\textbf {\bibinfo {volume} {14}},\ \bibinfo {pages} {5195} (\bibinfo {year} {1981})}\BibitemShut {NoStop}%
\bibitem [{\citenamefont {Nataf}\ \emph {et~al.}(2020)\citenamefont {Nataf}, \citenamefont {Guennou}, \citenamefont {Gregg}, \citenamefont {Meier}, \citenamefont {Hlinka}, \citenamefont {Salje},\ and\ \citenamefont {Kreisel}}]{nataf2020domain}%
  \BibitemOpen
  \bibfield  {author} {\bibinfo {author} {\bibfnamefont {G.}~\bibnamefont {Nataf}}, \bibinfo {author} {\bibfnamefont {M.}~\bibnamefont {Guennou}}, \bibinfo {author} {\bibfnamefont {J.}~\bibnamefont {Gregg}}, \bibinfo {author} {\bibfnamefont {D.}~\bibnamefont {Meier}}, \bibinfo {author} {\bibfnamefont {J.}~\bibnamefont {Hlinka}}, \bibinfo {author} {\bibfnamefont {E.}~\bibnamefont {Salje}},\ and\ \bibinfo {author} {\bibfnamefont {J.}~\bibnamefont {Kreisel}},\ }\bibfield  {title} {\bibinfo {title} {Domain-wall engineering and topological defects in ferroelectric and ferroelastic materials},\ }\href@noop {} {\bibfield  {journal} {\bibinfo  {journal} {Nature Reviews Physics}\ }\textbf {\bibinfo {volume} {2}},\ \bibinfo {pages} {634} (\bibinfo {year} {2020})}\BibitemShut {NoStop}%
\bibitem [{\citenamefont {Menath}\ \emph {et~al.}(2023)\citenamefont {Menath}, \citenamefont {Mohammadi}, \citenamefont {Grauer}, \citenamefont {Deters}, \citenamefont {B{\"o}hm}, \citenamefont {Liebchen}, \citenamefont {Janssen}, \citenamefont {L{\"o}wen},\ and\ \citenamefont {Vogel}}]{menath2023acoustic}%
  \BibitemOpen
  \bibfield  {author} {\bibinfo {author} {\bibfnamefont {J.}~\bibnamefont {Menath}}, \bibinfo {author} {\bibfnamefont {R.}~\bibnamefont {Mohammadi}}, \bibinfo {author} {\bibfnamefont {J.~C.}\ \bibnamefont {Grauer}}, \bibinfo {author} {\bibfnamefont {C.}~\bibnamefont {Deters}}, \bibinfo {author} {\bibfnamefont {M.}~\bibnamefont {B{\"o}hm}}, \bibinfo {author} {\bibfnamefont {B.}~\bibnamefont {Liebchen}}, \bibinfo {author} {\bibfnamefont {L.~M.}\ \bibnamefont {Janssen}}, \bibinfo {author} {\bibfnamefont {H.}~\bibnamefont {L{\"o}wen}},\ and\ \bibinfo {author} {\bibfnamefont {N.}~\bibnamefont {Vogel}},\ }\bibfield  {title} {\bibinfo {title} {Acoustic crystallization of 2d colloidal crystals},\ }\href@noop {} {\bibfield  {journal} {\bibinfo  {journal} {Advanced Materials}\ }\textbf {\bibinfo {volume} {35}},\ \bibinfo {pages} {2206593} (\bibinfo {year} {2023})}\BibitemShut {NoStop}%
\bibitem [{\citenamefont {He}\ \emph {et~al.}(2024)\citenamefont {He}, \citenamefont {Qiu}, \citenamefont {Wang}, \citenamefont {Jin}, \citenamefont {Zhou}, \citenamefont {Giersig}, \citenamefont {Kempa},\ and\ \citenamefont {Akinoglu}}]{he2024regulating}%
  \BibitemOpen
  \bibfield  {author} {\bibinfo {author} {\bibfnamefont {G.}~\bibnamefont {He}}, \bibinfo {author} {\bibfnamefont {T.}~\bibnamefont {Qiu}}, \bibinfo {author} {\bibfnamefont {X.}~\bibnamefont {Wang}}, \bibinfo {author} {\bibfnamefont {M.}~\bibnamefont {Jin}}, \bibinfo {author} {\bibfnamefont {G.}~\bibnamefont {Zhou}}, \bibinfo {author} {\bibfnamefont {M.}~\bibnamefont {Giersig}}, \bibinfo {author} {\bibfnamefont {K.}~\bibnamefont {Kempa}},\ and\ \bibinfo {author} {\bibfnamefont {E.~M.}\ \bibnamefont {Akinoglu}},\ }\bibfield  {title} {\bibinfo {title} {Regulating two-dimensional colloidal crystal assembly through contactless acoustic annealing},\ }\href@noop {} {\bibfield  {journal} {\bibinfo  {journal} {Journal of Applied Physics}\ }\textbf {\bibinfo {volume} {135}} (\bibinfo {year} {2024})}\BibitemShut {NoStop}%
\bibitem [{\citenamefont {Chen}\ \emph {et~al.}(2014)\citenamefont {Chen}, \citenamefont {Upadhyaya},\ and\ \citenamefont {Vitelli}}]{chen2014nonlinear}%
  \BibitemOpen
  \bibfield  {author} {\bibinfo {author} {\bibfnamefont {B.~G.-g.}\ \bibnamefont {Chen}}, \bibinfo {author} {\bibfnamefont {N.}~\bibnamefont {Upadhyaya}},\ and\ \bibinfo {author} {\bibfnamefont {V.}~\bibnamefont {Vitelli}},\ }\bibfield  {title} {\bibinfo {title} {Nonlinear conduction via solitons in a topological mechanical insulator},\ }\href@noop {} {\bibfield  {journal} {\bibinfo  {journal} {Proceedings of the National Academy of Sciences}\ }\textbf {\bibinfo {volume} {111}},\ \bibinfo {pages} {13004} (\bibinfo {year} {2014})}\BibitemShut {NoStop}%
\bibitem [{\citenamefont {Zhou}\ \emph {et~al.}(2017)\citenamefont {Zhou}, \citenamefont {Chen}, \citenamefont {Upadhyaya}, \citenamefont {Vitelli} \emph {et~al.}}]{zhou2017kink}%
  \BibitemOpen
  \bibfield  {author} {\bibinfo {author} {\bibfnamefont {Y.}~\bibnamefont {Zhou}}, \bibinfo {author} {\bibfnamefont {B.~G.-g.}\ \bibnamefont {Chen}}, \bibinfo {author} {\bibfnamefont {N.}~\bibnamefont {Upadhyaya}}, \bibinfo {author} {\bibfnamefont {V.}~\bibnamefont {Vitelli}}, \emph {et~al.},\ }\bibfield  {title} {\bibinfo {title} {Kink-antikink asymmetry and impurity interactions in topological mechanical chains},\ }\href@noop {} {\bibfield  {journal} {\bibinfo  {journal} {Physical Review E}\ }\textbf {\bibinfo {volume} {95}},\ \bibinfo {pages} {022202} (\bibinfo {year} {2017})}\BibitemShut {NoStop}%
\bibitem [{\citenamefont {Mao}\ and\ \citenamefont {Lubensky}(2018)}]{mao2018maxwell}%
  \BibitemOpen
  \bibfield  {author} {\bibinfo {author} {\bibfnamefont {X.}~\bibnamefont {Mao}}\ and\ \bibinfo {author} {\bibfnamefont {T.~C.}\ \bibnamefont {Lubensky}},\ }\bibfield  {title} {\bibinfo {title} {Maxwell lattices and topological mechanics},\ }\href@noop {} {\bibfield  {journal} {\bibinfo  {journal} {Annual Review of Condensed Matter Physics}\ }\textbf {\bibinfo {volume} {9}},\ \bibinfo {pages} {413} (\bibinfo {year} {2018})}\BibitemShut {NoStop}%
\bibitem [{\citenamefont {Zhang}\ \emph {et~al.}(2019)\citenamefont {Zhang}, \citenamefont {Li}, \citenamefont {Zheng}, \citenamefont {Genin},\ and\ \citenamefont {Chen}}]{zhang2019programmable}%
  \BibitemOpen
  \bibfield  {author} {\bibinfo {author} {\bibfnamefont {Y.}~\bibnamefont {Zhang}}, \bibinfo {author} {\bibfnamefont {B.}~\bibnamefont {Li}}, \bibinfo {author} {\bibfnamefont {Q.}~\bibnamefont {Zheng}}, \bibinfo {author} {\bibfnamefont {G.~M.}\ \bibnamefont {Genin}},\ and\ \bibinfo {author} {\bibfnamefont {C.}~\bibnamefont {Chen}},\ }\bibfield  {title} {\bibinfo {title} {Programmable and robust static topological solitons in mechanical metamaterials},\ }\href@noop {} {\bibfield  {journal} {\bibinfo  {journal} {Nature communications}\ }\textbf {\bibinfo {volume} {10}},\ \bibinfo {pages} {5605} (\bibinfo {year} {2019})}\BibitemShut {NoStop}%
\bibitem [{\citenamefont {Deng}\ \emph {et~al.}(2020)\citenamefont {Deng}, \citenamefont {Yu}, \citenamefont {Forte}, \citenamefont {Tournat},\ and\ \citenamefont {Bertoldi}}]{deng2020characterization}%
  \BibitemOpen
  \bibfield  {author} {\bibinfo {author} {\bibfnamefont {B.}~\bibnamefont {Deng}}, \bibinfo {author} {\bibfnamefont {S.}~\bibnamefont {Yu}}, \bibinfo {author} {\bibfnamefont {A.~E.}\ \bibnamefont {Forte}}, \bibinfo {author} {\bibfnamefont {V.}~\bibnamefont {Tournat}},\ and\ \bibinfo {author} {\bibfnamefont {K.}~\bibnamefont {Bertoldi}},\ }\bibfield  {title} {\bibinfo {title} {Characterization, stability, and application of domain walls in flexible mechanical metamaterials},\ }\href@noop {} {\bibfield  {journal} {\bibinfo  {journal} {Proceedings of the National Academy of Sciences}\ }\textbf {\bibinfo {volume} {117}},\ \bibinfo {pages} {31002} (\bibinfo {year} {2020})}\BibitemShut {NoStop}%
\bibitem [{\citenamefont {Upadhyaya}\ \emph {et~al.}(2020)\citenamefont {Upadhyaya}, \citenamefont {Chen},\ and\ \citenamefont {Vitelli}}]{upadhyaya2020nuts}%
  \BibitemOpen
  \bibfield  {author} {\bibinfo {author} {\bibfnamefont {N.}~\bibnamefont {Upadhyaya}}, \bibinfo {author} {\bibfnamefont {B.~G.}\ \bibnamefont {Chen}},\ and\ \bibinfo {author} {\bibfnamefont {V.}~\bibnamefont {Vitelli}},\ }\bibfield  {title} {\bibinfo {title} {Nuts and bolts of supersymmetry},\ }\href@noop {} {\bibfield  {journal} {\bibinfo  {journal} {Physical Review Research}\ }\textbf {\bibinfo {volume} {2}},\ \bibinfo {pages} {043098} (\bibinfo {year} {2020})}\BibitemShut {NoStop}%
\bibitem [{\citenamefont {Lo}\ \emph {et~al.}(2021)\citenamefont {Lo}, \citenamefont {Santangelo}, \citenamefont {Chen}, \citenamefont {Jian}, \citenamefont {Roychowdhury},\ and\ \citenamefont {Lawler}}]{lo2021topology}%
  \BibitemOpen
  \bibfield  {author} {\bibinfo {author} {\bibfnamefont {P.-W.}\ \bibnamefont {Lo}}, \bibinfo {author} {\bibfnamefont {C.~D.}\ \bibnamefont {Santangelo}}, \bibinfo {author} {\bibfnamefont {B.~G.-g.}\ \bibnamefont {Chen}}, \bibinfo {author} {\bibfnamefont {C.-M.}\ \bibnamefont {Jian}}, \bibinfo {author} {\bibfnamefont {K.}~\bibnamefont {Roychowdhury}},\ and\ \bibinfo {author} {\bibfnamefont {M.~J.}\ \bibnamefont {Lawler}},\ }\bibfield  {title} {\bibinfo {title} {Topology in nonlinear mechanical systems},\ }\href@noop {} {\bibfield  {journal} {\bibinfo  {journal} {Physical Review Letters}\ }\textbf {\bibinfo {volume} {127}},\ \bibinfo {pages} {076802} (\bibinfo {year} {2021})}\BibitemShut {NoStop}%
\bibitem [{\citenamefont {Zhou}\ \emph {et~al.}(2021)\citenamefont {Zhou}, \citenamefont {Zhang},\ and\ \citenamefont {Chen}}]{zhou2021amplitude}%
  \BibitemOpen
  \bibfield  {author} {\bibinfo {author} {\bibfnamefont {Y.}~\bibnamefont {Zhou}}, \bibinfo {author} {\bibfnamefont {Y.}~\bibnamefont {Zhang}},\ and\ \bibinfo {author} {\bibfnamefont {C.}~\bibnamefont {Chen}},\ }\bibfield  {title} {\bibinfo {title} {Amplitude-dependent boundary modes in topological mechanical lattices},\ }\href@noop {} {\bibfield  {journal} {\bibinfo  {journal} {Journal of the Mechanics and Physics of Solids}\ }\textbf {\bibinfo {volume} {153}},\ \bibinfo {pages} {104482} (\bibinfo {year} {2021})}\BibitemShut {NoStop}%
\bibitem [{\citenamefont {Sun}\ and\ \citenamefont {Mao}(2021)}]{sun2021fractional}%
  \BibitemOpen
  \bibfield  {author} {\bibinfo {author} {\bibfnamefont {K.}~\bibnamefont {Sun}}\ and\ \bibinfo {author} {\bibfnamefont {X.}~\bibnamefont {Mao}},\ }\bibfield  {title} {\bibinfo {title} {Fractional excitations in non-euclidean elastic plates},\ }\href@noop {} {\bibfield  {journal} {\bibinfo  {journal} {Physical Review Letters}\ }\textbf {\bibinfo {volume} {127}},\ \bibinfo {pages} {098001} (\bibinfo {year} {2021})}\BibitemShut {NoStop}%
\bibitem [{\citenamefont {Deng}\ \emph {et~al.}(2022)\citenamefont {Deng}, \citenamefont {Zanaty}, \citenamefont {Forte},\ and\ \citenamefont {Bertoldi}}]{deng2022topological}%
  \BibitemOpen
  \bibfield  {author} {\bibinfo {author} {\bibfnamefont {B.}~\bibnamefont {Deng}}, \bibinfo {author} {\bibfnamefont {M.}~\bibnamefont {Zanaty}}, \bibinfo {author} {\bibfnamefont {A.~E.}\ \bibnamefont {Forte}},\ and\ \bibinfo {author} {\bibfnamefont {K.}~\bibnamefont {Bertoldi}},\ }\bibfield  {title} {\bibinfo {title} {Topological solitons make metamaterials crawl},\ }\href@noop {} {\bibfield  {journal} {\bibinfo  {journal} {Physical Review Applied}\ }\textbf {\bibinfo {volume} {17}},\ \bibinfo {pages} {014004} (\bibinfo {year} {2022})}\BibitemShut {NoStop}%
\bibitem [{\citenamefont {N{\'u}{\~n}ez}\ \emph {et~al.}(2023)\citenamefont {N{\'u}{\~n}ez}, \citenamefont {Poli}, \citenamefont {Stanifer}, \citenamefont {Mao},\ and\ \citenamefont {Arruda}}]{nunez2023fractional}%
  \BibitemOpen
  \bibfield  {author} {\bibinfo {author} {\bibfnamefont {C.~N.~V.}\ \bibnamefont {N{\'u}{\~n}ez}}, \bibinfo {author} {\bibfnamefont {A.}~\bibnamefont {Poli}}, \bibinfo {author} {\bibfnamefont {E.}~\bibnamefont {Stanifer}}, \bibinfo {author} {\bibfnamefont {X.}~\bibnamefont {Mao}},\ and\ \bibinfo {author} {\bibfnamefont {E.~M.}\ \bibnamefont {Arruda}},\ }\bibfield  {title} {\bibinfo {title} {Fractional topological solitons in nonlinear viscoelastic ribbons with tunable speed},\ }\href@noop {} {\bibfield  {journal} {\bibinfo  {journal} {Extreme Mechanics Letters}\ }\textbf {\bibinfo {volume} {61}},\ \bibinfo {pages} {102027} (\bibinfo {year} {2023})}\BibitemShut {NoStop}%
\bibitem [{\citenamefont {Zhou}\ \emph {et~al.}(2024)\citenamefont {Zhou}, \citenamefont {Zhang}, \citenamefont {Long}, \citenamefont {Wang},\ and\ \citenamefont {Chen}}]{zhou2024static}%
  \BibitemOpen
  \bibfield  {author} {\bibinfo {author} {\bibfnamefont {Y.}~\bibnamefont {Zhou}}, \bibinfo {author} {\bibfnamefont {Y.}~\bibnamefont {Zhang}}, \bibinfo {author} {\bibfnamefont {J.}~\bibnamefont {Long}}, \bibinfo {author} {\bibfnamefont {A.}~\bibnamefont {Wang}},\ and\ \bibinfo {author} {\bibfnamefont {C.~Q.}\ \bibnamefont {Chen}},\ }\bibfield  {title} {\bibinfo {title} {Static vector solitons in a topological mechanical lattice},\ }\href@noop {} {\bibfield  {journal} {\bibinfo  {journal} {Communications Physics}\ }\textbf {\bibinfo {volume} {7}},\ \bibinfo {pages} {131} (\bibinfo {year} {2024})}\BibitemShut {NoStop}%
\bibitem [{\citenamefont {Gorman}\ \emph {et~al.}(1969)\citenamefont {Gorman}, \citenamefont {Wood},\ and\ \citenamefont {Vreeland~Jr}}]{gorman1969mobility}%
  \BibitemOpen
  \bibfield  {author} {\bibinfo {author} {\bibfnamefont {J.}~\bibnamefont {Gorman}}, \bibinfo {author} {\bibfnamefont {D.}~\bibnamefont {Wood}},\ and\ \bibinfo {author} {\bibfnamefont {T.}~\bibnamefont {Vreeland~Jr}},\ }\bibfield  {title} {\bibinfo {title} {Mobility of dislocations in aluminum},\ }\href@noop {} {\bibfield  {journal} {\bibinfo  {journal} {Journal of Applied Physics}\ }\textbf {\bibinfo {volume} {40}},\ \bibinfo {pages} {833} (\bibinfo {year} {1969})}\BibitemShut {NoStop}%
\bibitem [{\citenamefont {Gremaud}\ \emph {et~al.}(1987)\citenamefont {Gremaud}, \citenamefont {Bujard},\ and\ \citenamefont {Benoit}}]{gremaud1987coupling}%
  \BibitemOpen
  \bibfield  {author} {\bibinfo {author} {\bibfnamefont {G.}~\bibnamefont {Gremaud}}, \bibinfo {author} {\bibfnamefont {M.}~\bibnamefont {Bujard}},\ and\ \bibinfo {author} {\bibfnamefont {W.}~\bibnamefont {Benoit}},\ }\bibfield  {title} {\bibinfo {title} {The coupling technique: A two-wave acoustic method for the study of dislocation dynamics},\ }\href@noop {} {\bibfield  {journal} {\bibinfo  {journal} {Journal of applied physics}\ }\textbf {\bibinfo {volume} {61}},\ \bibinfo {pages} {1795} (\bibinfo {year} {1987})}\BibitemShut {NoStop}%
\bibitem [{\citenamefont {Conrad}(1964)}]{conrad1964thermally}%
  \BibitemOpen
  \bibfield  {author} {\bibinfo {author} {\bibfnamefont {H.}~\bibnamefont {Conrad}},\ }\bibfield  {title} {\bibinfo {title} {Thermally activated deformation of metals},\ }\href@noop {} {\bibfield  {journal} {\bibinfo  {journal} {Jom}\ }\textbf {\bibinfo {volume} {16}},\ \bibinfo {pages} {582} (\bibinfo {year} {1964})}\BibitemShut {NoStop}%
\bibitem [{\citenamefont {Hutchison}\ \emph {et~al.}(1965)\citenamefont {Hutchison}, \citenamefont {Rogers},\ and\ \citenamefont {Turkington}}]{hutchison1965thermally}%
  \BibitemOpen
  \bibfield  {author} {\bibinfo {author} {\bibfnamefont {T.}~\bibnamefont {Hutchison}}, \bibinfo {author} {\bibfnamefont {D.}~\bibnamefont {Rogers}},\ and\ \bibinfo {author} {\bibfnamefont {R.}~\bibnamefont {Turkington}},\ }\bibfield  {title} {\bibinfo {title} {Thermally activated dislocation motion in aluminum},\ }\href@noop {} {\bibfield  {journal} {\bibinfo  {journal} {Journal of Applied Physics}\ }\textbf {\bibinfo {volume} {36}},\ \bibinfo {pages} {871} (\bibinfo {year} {1965})}\BibitemShut {NoStop}%
\bibitem [{\citenamefont {Ogata}\ \emph {et~al.}(1986)\citenamefont {Ogata}, \citenamefont {Terai},\ and\ \citenamefont {Wada}}]{ogata1986brownian}%
  \BibitemOpen
  \bibfield  {author} {\bibinfo {author} {\bibfnamefont {M.}~\bibnamefont {Ogata}}, \bibinfo {author} {\bibfnamefont {A.}~\bibnamefont {Terai}},\ and\ \bibinfo {author} {\bibfnamefont {Y.}~\bibnamefont {Wada}},\ }\bibfield  {title} {\bibinfo {title} {Brownian motion of a soliton in trans-polyacetylene. i. random walk mechanism},\ }\href@noop {} {\bibfield  {journal} {\bibinfo  {journal} {Journal of the Physical Society of Japan}\ }\textbf {\bibinfo {volume} {55}},\ \bibinfo {pages} {2305} (\bibinfo {year} {1986})}\BibitemShut {NoStop}%
\bibitem [{Note1()}]{Note1}%
  \BibitemOpen
  \bibinfo {note} {Another, albeit less, related area is phonon interplay with multistable systems supporting transition waves (\protect \textit {e.g.}, Ref.~\cite {khomeriki2008tristability, wu2018metastable, hwang2021extreme}). We note that these systems do not support kinks as they do not have symmetric ground states and use high-amplitude excitation to initiate transitions}\BibitemShut {NoStop}%
\bibitem [{\citenamefont {Hasenfratz}\ and\ \citenamefont {Klein}(1977)}]{hasenfratz1977interaction}%
  \BibitemOpen
  \bibfield  {author} {\bibinfo {author} {\bibfnamefont {W.}~\bibnamefont {Hasenfratz}}\ and\ \bibinfo {author} {\bibfnamefont {R.}~\bibnamefont {Klein}},\ }\bibfield  {title} {\bibinfo {title} {The interaction of a solitary wave solution with phonons in a one-dimensional model for displacive structural phase transitions},\ }\href@noop {} {\bibfield  {journal} {\bibinfo  {journal} {Physica A: Statistical Mechanics and its Applications}\ }\textbf {\bibinfo {volume} {89}},\ \bibinfo {pages} {191} (\bibinfo {year} {1977})}\BibitemShut {NoStop}%
\bibitem [{\citenamefont {Wada}\ and\ \citenamefont {Schrieffer}(1978)}]{wada1978brownian}%
  \BibitemOpen
  \bibfield  {author} {\bibinfo {author} {\bibfnamefont {Y.}~\bibnamefont {Wada}}\ and\ \bibinfo {author} {\bibfnamefont {J.}~\bibnamefont {Schrieffer}},\ }\bibfield  {title} {\bibinfo {title} {Brownian motion of a domain wall and the diffusion constants},\ }\href@noop {} {\bibfield  {journal} {\bibinfo  {journal} {Physical Review B}\ }\textbf {\bibinfo {volume} {18}},\ \bibinfo {pages} {3897} (\bibinfo {year} {1978})}\BibitemShut {NoStop}%
\bibitem [{\citenamefont {Theodorakopoulos}(1979)}]{theodorakopoulos1979dynamics}%
  \BibitemOpen
  \bibfield  {author} {\bibinfo {author} {\bibfnamefont {N.}~\bibnamefont {Theodorakopoulos}},\ }\bibfield  {title} {\bibinfo {title} {Dynamics of non-linear systems: The kink-phonon interaction},\ }\href@noop {} {\bibfield  {journal} {\bibinfo  {journal} {Zeitschrift f{\"u}r Physik B Condensed Matter}\ }\textbf {\bibinfo {volume} {33}},\ \bibinfo {pages} {385} (\bibinfo {year} {1979})}\BibitemShut {NoStop}%
\bibitem [{\citenamefont {Theodorakopoulos}\ \emph {et~al.}(1980)\citenamefont {Theodorakopoulos}, \citenamefont {W{\"u}nderlich},\ and\ \citenamefont {Klein}}]{theodorakopoulos1980lattice}%
  \BibitemOpen
  \bibfield  {author} {\bibinfo {author} {\bibfnamefont {N.}~\bibnamefont {Theodorakopoulos}}, \bibinfo {author} {\bibfnamefont {W.}~\bibnamefont {W{\"u}nderlich}},\ and\ \bibinfo {author} {\bibfnamefont {R.}~\bibnamefont {Klein}},\ }\bibfield  {title} {\bibinfo {title} {Lattice phonons in the presence of non-linear excitations},\ }\href@noop {} {\bibfield  {journal} {\bibinfo  {journal} {Solid State Communications}\ }\textbf {\bibinfo {volume} {33}},\ \bibinfo {pages} {213} (\bibinfo {year} {1980})}\BibitemShut {NoStop}%
\bibitem [{\citenamefont {Klein}\ \emph {et~al.}(1980)\citenamefont {Klein}, \citenamefont {Hasenfratz}, \citenamefont {Theodorakopoulos},\ and\ \citenamefont {W{\"u}nderlich}}]{klein1980kink}%
  \BibitemOpen
  \bibfield  {author} {\bibinfo {author} {\bibfnamefont {R.}~\bibnamefont {Klein}}, \bibinfo {author} {\bibfnamefont {W.}~\bibnamefont {Hasenfratz}}, \bibinfo {author} {\bibfnamefont {N.}~\bibnamefont {Theodorakopoulos}},\ and\ \bibinfo {author} {\bibfnamefont {W.}~\bibnamefont {W{\"u}nderlich}},\ }\bibfield  {title} {\bibinfo {title} {The kink-phonon and the kink-kink interaction in the $\phi$4 model.},\ }\href@noop {} {\bibfield  {journal} {\bibinfo  {journal} {Ferroelectrics}\ }\textbf {\bibinfo {volume} {26}},\ \bibinfo {pages} {721} (\bibinfo {year} {1980})}\BibitemShut {NoStop}%
\bibitem [{\citenamefont {Ishiuchi}\ and\ \citenamefont {Wada}(1980)}]{ishiuchi1980brownian}%
  \BibitemOpen
  \bibfield  {author} {\bibinfo {author} {\bibfnamefont {H.}~\bibnamefont {Ishiuchi}}\ and\ \bibinfo {author} {\bibfnamefont {Y.}~\bibnamefont {Wada}},\ }\bibfield  {title} {\bibinfo {title} {Brownian motion of a domain wall with higher order phonon interactions},\ }\href@noop {} {\bibfield  {journal} {\bibinfo  {journal} {Progress of Theoretical Physics Supplement}\ }\textbf {\bibinfo {volume} {69}},\ \bibinfo {pages} {242} (\bibinfo {year} {1980})}\BibitemShut {NoStop}%
\bibitem [{\citenamefont {Ogata}\ and\ \citenamefont {Wada}(1984)}]{ogata1984momentum}%
  \BibitemOpen
  \bibfield  {author} {\bibinfo {author} {\bibfnamefont {M.}~\bibnamefont {Ogata}}\ and\ \bibinfo {author} {\bibfnamefont {Y.}~\bibnamefont {Wada}},\ }\bibfield  {title} {\bibinfo {title} {Momentum transfer between a kink and a phonon in the one-dimensional $\varphi$4 system},\ }\href@noop {} {\bibfield  {journal} {\bibinfo  {journal} {Journal of the Physical Society of Japan}\ }\textbf {\bibinfo {volume} {53}},\ \bibinfo {pages} {3855} (\bibinfo {year} {1984})}\BibitemShut {NoStop}%
\bibitem [{\citenamefont {Abdelhady}\ and\ \citenamefont {Weigel}(2011)}]{abdelhady2011wave}%
  \BibitemOpen
  \bibfield  {author} {\bibinfo {author} {\bibfnamefont {A.}~\bibnamefont {Abdelhady}}\ and\ \bibinfo {author} {\bibfnamefont {H.}~\bibnamefont {Weigel}},\ }\bibfield  {title} {\bibinfo {title} {Wave-packet scattering off the kink-solution},\ }\href@noop {} {\bibfield  {journal} {\bibinfo  {journal} {International Journal of Modern Physics A}\ }\textbf {\bibinfo {volume} {26}},\ \bibinfo {pages} {3625} (\bibinfo {year} {2011})}\BibitemShut {NoStop}%
\bibitem [{\citenamefont {Theodorakopoulos}(1980)}]{theodorakopoulos1980dynamics}%
  \BibitemOpen
  \bibfield  {author} {\bibinfo {author} {\bibfnamefont {N.}~\bibnamefont {Theodorakopoulos}},\ }\bibfield  {title} {\bibinfo {title} {Dynamics of the sine-gordon chain: The kink-phonon interaction, soliton diffusion and dynamical correlations},\ }in\ \href@noop {} {\emph {\bibinfo {booktitle} {Ordering in Strongly Fluctuating Condensed Matter Systems}}}\ (\bibinfo  {publisher} {Springer},\ \bibinfo {year} {1980})\ pp.\ \bibinfo {pages} {145--149}\BibitemShut {NoStop}%
\bibitem [{\citenamefont {Wada}\ and\ \citenamefont {Ishiuchi}(1982)}]{wada1982brownian}%
  \BibitemOpen
  \bibfield  {author} {\bibinfo {author} {\bibfnamefont {Y.}~\bibnamefont {Wada}}\ and\ \bibinfo {author} {\bibfnamefont {H.}~\bibnamefont {Ishiuchi}},\ }\bibfield  {title} {\bibinfo {title} {Brownian motion of a kink in sine-gordon system and diffusion constant},\ }\href@noop {} {\bibfield  {journal} {\bibinfo  {journal} {Journal of the Physical Society of Japan}\ }\textbf {\bibinfo {volume} {51}},\ \bibinfo {pages} {1372} (\bibinfo {year} {1982})}\BibitemShut {NoStop}%
\bibitem [{\citenamefont {Peierls}(1940)}]{peierls1940size}%
  \BibitemOpen
  \bibfield  {author} {\bibinfo {author} {\bibfnamefont {R.}~\bibnamefont {Peierls}},\ }\bibfield  {title} {\bibinfo {title} {The size of a dislocation},\ }\href@noop {} {\bibfield  {journal} {\bibinfo  {journal} {Proceedings of the Physical Society}\ }\textbf {\bibinfo {volume} {52}},\ \bibinfo {pages} {34} (\bibinfo {year} {1940})}\BibitemShut {NoStop}%
\bibitem [{\citenamefont {Nabarro}(1947)}]{nabarro1947dislocations}%
  \BibitemOpen
  \bibfield  {author} {\bibinfo {author} {\bibfnamefont {F.}~\bibnamefont {Nabarro}},\ }\bibfield  {title} {\bibinfo {title} {Dislocations in a simple cubic lattice},\ }\href@noop {} {\bibfield  {journal} {\bibinfo  {journal} {Proceedings of the Physical Society}\ }\textbf {\bibinfo {volume} {59}},\ \bibinfo {pages} {256} (\bibinfo {year} {1947})}\BibitemShut {NoStop}%
\bibitem [{\citenamefont {Kivshar}\ and\ \citenamefont {Campbell}(1993)}]{kivshar1993peierls}%
  \BibitemOpen
  \bibfield  {author} {\bibinfo {author} {\bibfnamefont {Y.~S.}\ \bibnamefont {Kivshar}}\ and\ \bibinfo {author} {\bibfnamefont {D.~K.}\ \bibnamefont {Campbell}},\ }\bibfield  {title} {\bibinfo {title} {Peierls-nabarro potential barrier for highly localized nonlinear modes},\ }\href@noop {} {\bibfield  {journal} {\bibinfo  {journal} {Physical Review E}\ }\textbf {\bibinfo {volume} {48}},\ \bibinfo {pages} {3077} (\bibinfo {year} {1993})}\BibitemShut {NoStop}%
\bibitem [{\citenamefont {Braun}\ and\ \citenamefont {Kivshar}(2004)}]{braun2004frenkel}%
  \BibitemOpen
  \bibfield  {author} {\bibinfo {author} {\bibfnamefont {O.~M.}\ \bibnamefont {Braun}}\ and\ \bibinfo {author} {\bibfnamefont {Y.~S.}\ \bibnamefont {Kivshar}},\ }\href@noop {} {\emph {\bibinfo {title} {The Frenkel-Kontorova model: concepts, methods, and applications}}},\ Vol.~\bibinfo {volume} {18}\ (\bibinfo  {publisher} {Springer},\ \bibinfo {year} {2004})\BibitemShut {NoStop}%
\bibitem [{\citenamefont {Chirilus-Bruckner}\ \emph {et~al.}(2014)\citenamefont {Chirilus-Bruckner}, \citenamefont {Chong}, \citenamefont {Cuevas-Maraver},\ and\ \citenamefont {Kevrekidis}}]{chirilus2014sine}%
  \BibitemOpen
  \bibfield  {author} {\bibinfo {author} {\bibfnamefont {M.}~\bibnamefont {Chirilus-Bruckner}}, \bibinfo {author} {\bibfnamefont {C.}~\bibnamefont {Chong}}, \bibinfo {author} {\bibfnamefont {J.}~\bibnamefont {Cuevas-Maraver}},\ and\ \bibinfo {author} {\bibfnamefont {P.}~\bibnamefont {Kevrekidis}},\ }\bibfield  {title} {\bibinfo {title} {Sine-gordon equation: From discrete to continuum},\ }\href@noop {} {\bibfield  {journal} {\bibinfo  {journal} {The sine-Gordon Model and its Applications: From Pendula and Josephson Junctions to Gravity and High-Energy Physics}\ ,\ \bibinfo {pages} {31}} (\bibinfo {year} {2014})}\BibitemShut {NoStop}%
\bibitem [{\citenamefont {Kevrekidis}\ and\ \citenamefont {Cuevas-Maraver}(2019)}]{kevrekidis2019dynamical}%
  \BibitemOpen
  \bibfield  {author} {\bibinfo {author} {\bibfnamefont {P.~G.}\ \bibnamefont {Kevrekidis}}\ and\ \bibinfo {author} {\bibfnamefont {J.}~\bibnamefont {Cuevas-Maraver}},\ }\bibfield  {title} {\bibinfo {title} {A dynamical perspective on the $\varphi$4 model},\ }\href@noop {} {\bibfield  {journal} {\bibinfo  {journal} {Past, Present and Future}\ }\textbf {\bibinfo {volume} {26}} (\bibinfo {year} {2019})}\BibitemShut {NoStop}%
\bibitem [{\citenamefont {Peyrard}\ and\ \citenamefont {Kruskal}(1984)}]{peyrard1984kink}%
  \BibitemOpen
  \bibfield  {author} {\bibinfo {author} {\bibfnamefont {M.}~\bibnamefont {Peyrard}}\ and\ \bibinfo {author} {\bibfnamefont {M.~D.}\ \bibnamefont {Kruskal}},\ }\bibfield  {title} {\bibinfo {title} {Kink dynamics in the highly discrete sine-gordon system},\ }\href@noop {} {\bibfield  {journal} {\bibinfo  {journal} {Physica D: Nonlinear Phenomena}\ }\textbf {\bibinfo {volume} {14}},\ \bibinfo {pages} {88} (\bibinfo {year} {1984})}\BibitemShut {NoStop}%
\bibitem [{\citenamefont {Speight}\ and\ \citenamefont {Ward}(1994)}]{speight1994kink}%
  \BibitemOpen
  \bibfield  {author} {\bibinfo {author} {\bibfnamefont {J.}~\bibnamefont {Speight}}\ and\ \bibinfo {author} {\bibfnamefont {R.}~\bibnamefont {Ward}},\ }\bibfield  {title} {\bibinfo {title} {Kink dynamics in a novel discrete sine-gordon system},\ }\href@noop {} {\bibfield  {journal} {\bibinfo  {journal} {Nonlinearity}\ }\textbf {\bibinfo {volume} {7}},\ \bibinfo {pages} {475} (\bibinfo {year} {1994})}\BibitemShut {NoStop}%
\bibitem [{\citenamefont {Speight}(1997)}]{speight1997discrete}%
  \BibitemOpen
  \bibfield  {author} {\bibinfo {author} {\bibfnamefont {J.}~\bibnamefont {Speight}},\ }\bibfield  {title} {\bibinfo {title} {A discrete system without a peierls-nabarro barrier},\ }\href@noop {} {\bibfield  {journal} {\bibinfo  {journal} {Nonlinearity}\ }\textbf {\bibinfo {volume} {10}},\ \bibinfo {pages} {1615} (\bibinfo {year} {1997})}\BibitemShut {NoStop}%
\bibitem [{\citenamefont {Speight}(1999)}]{speight1999topological}%
  \BibitemOpen
  \bibfield  {author} {\bibinfo {author} {\bibfnamefont {J.~M.}\ \bibnamefont {Speight}},\ }\bibfield  {title} {\bibinfo {title} {Topological discrete kinks},\ }\href@noop {} {\bibfield  {journal} {\bibinfo  {journal} {Nonlinearity}\ }\textbf {\bibinfo {volume} {12}},\ \bibinfo {pages} {1373} (\bibinfo {year} {1999})}\BibitemShut {NoStop}%
\bibitem [{\citenamefont {Flach}\ \emph {et~al.}(1999)\citenamefont {Flach}, \citenamefont {Zolotaryuk},\ and\ \citenamefont {Kladko}}]{flach1999moving}%
  \BibitemOpen
  \bibfield  {author} {\bibinfo {author} {\bibfnamefont {S.}~\bibnamefont {Flach}}, \bibinfo {author} {\bibfnamefont {Y.}~\bibnamefont {Zolotaryuk}},\ and\ \bibinfo {author} {\bibfnamefont {K.}~\bibnamefont {Kladko}},\ }\bibfield  {title} {\bibinfo {title} {Moving lattice kinks and pulses: An inverse method},\ }\href@noop {} {\bibfield  {journal} {\bibinfo  {journal} {Physical Review E}\ }\textbf {\bibinfo {volume} {59}},\ \bibinfo {pages} {6105} (\bibinfo {year} {1999})}\BibitemShut {NoStop}%
\bibitem [{\citenamefont {Savin}\ \emph {et~al.}(2000)\citenamefont {Savin}, \citenamefont {Zolotaryuk},\ and\ \citenamefont {Eilbeck}}]{savin2000moving}%
  \BibitemOpen
  \bibfield  {author} {\bibinfo {author} {\bibfnamefont {A.}~\bibnamefont {Savin}}, \bibinfo {author} {\bibfnamefont {Y.}~\bibnamefont {Zolotaryuk}},\ and\ \bibinfo {author} {\bibfnamefont {J.}~\bibnamefont {Eilbeck}},\ }\bibfield  {title} {\bibinfo {title} {Moving kinks and nanopterons in the nonlinear klein--gordon lattice},\ }\href@noop {} {\bibfield  {journal} {\bibinfo  {journal} {Physica D: Nonlinear Phenomena}\ }\textbf {\bibinfo {volume} {138}},\ \bibinfo {pages} {267} (\bibinfo {year} {2000})}\BibitemShut {NoStop}%
\bibitem [{\citenamefont {Kevrekidis}(2003)}]{kevrekidis2003class}%
  \BibitemOpen
  \bibfield  {author} {\bibinfo {author} {\bibfnamefont {P.}~\bibnamefont {Kevrekidis}},\ }\bibfield  {title} {\bibinfo {title} {On a class of discretizations of hamiltonian nonlinear partial differential equations},\ }\href@noop {} {\bibfield  {journal} {\bibinfo  {journal} {Physica D: Nonlinear Phenomena}\ }\textbf {\bibinfo {volume} {183}},\ \bibinfo {pages} {68} (\bibinfo {year} {2003})}\BibitemShut {NoStop}%
\bibitem [{\citenamefont {Aigner}\ \emph {et~al.}(2003)\citenamefont {Aigner}, \citenamefont {Champneys},\ and\ \citenamefont {Rothos}}]{aigner2003new}%
  \BibitemOpen
  \bibfield  {author} {\bibinfo {author} {\bibfnamefont {A.}~\bibnamefont {Aigner}}, \bibinfo {author} {\bibfnamefont {A.}~\bibnamefont {Champneys}},\ and\ \bibinfo {author} {\bibfnamefont {V.}~\bibnamefont {Rothos}},\ }\bibfield  {title} {\bibinfo {title} {A new barrier to the existence of moving kinks in frenkel--kontorova lattices},\ }\href@noop {} {\bibfield  {journal} {\bibinfo  {journal} {Physica D: Nonlinear Phenomena}\ }\textbf {\bibinfo {volume} {186}},\ \bibinfo {pages} {148} (\bibinfo {year} {2003})}\BibitemShut {NoStop}%
\bibitem [{\citenamefont {Cooper}\ \emph {et~al.}(2005)\citenamefont {Cooper}, \citenamefont {Khare}, \citenamefont {Mihaila},\ and\ \citenamefont {Saxena}}]{cooper2005exact}%
  \BibitemOpen
  \bibfield  {author} {\bibinfo {author} {\bibfnamefont {F.}~\bibnamefont {Cooper}}, \bibinfo {author} {\bibfnamefont {A.}~\bibnamefont {Khare}}, \bibinfo {author} {\bibfnamefont {B.}~\bibnamefont {Mihaila}},\ and\ \bibinfo {author} {\bibfnamefont {A.}~\bibnamefont {Saxena}},\ }\bibfield  {title} {\bibinfo {title} {Exact solitary wave solutions for a discrete $\lambda$ $\phi$ 4 field theory in 1+ 1 dimensions},\ }\href@noop {} {\bibfield  {journal} {\bibinfo  {journal} {Physical Review E—Statistical, Nonlinear, and Soft Matter Physics}\ }\textbf {\bibinfo {volume} {72}},\ \bibinfo {pages} {036605} (\bibinfo {year} {2005})}\BibitemShut {NoStop}%
\bibitem [{\citenamefont {Dmitriev}\ \emph {et~al.}(2005)\citenamefont {Dmitriev}, \citenamefont {Kevrekidis},\ and\ \citenamefont {Yoshikawa}}]{dmitriev2005discrete}%
  \BibitemOpen
  \bibfield  {author} {\bibinfo {author} {\bibfnamefont {S.}~\bibnamefont {Dmitriev}}, \bibinfo {author} {\bibfnamefont {P.}~\bibnamefont {Kevrekidis}},\ and\ \bibinfo {author} {\bibfnamefont {N.}~\bibnamefont {Yoshikawa}},\ }\bibfield  {title} {\bibinfo {title} {Discrete klein--gordon models with static kinks free of the peierls--nabarro potential},\ }\href@noop {} {\bibfield  {journal} {\bibinfo  {journal} {Journal of Physics A: Mathematical and General}\ }\textbf {\bibinfo {volume} {38}},\ \bibinfo {pages} {7617} (\bibinfo {year} {2005})}\BibitemShut {NoStop}%
\bibitem [{\citenamefont {Oxtoby}\ \emph {et~al.}(2005)\citenamefont {Oxtoby}, \citenamefont {Pelinovsky},\ and\ \citenamefont {Barashenkov}}]{oxtoby2005travelling}%
  \BibitemOpen
  \bibfield  {author} {\bibinfo {author} {\bibfnamefont {O.}~\bibnamefont {Oxtoby}}, \bibinfo {author} {\bibfnamefont {D.}~\bibnamefont {Pelinovsky}},\ and\ \bibinfo {author} {\bibfnamefont {I.}~\bibnamefont {Barashenkov}},\ }\bibfield  {title} {\bibinfo {title} {Travelling kinks in discrete $\phi$4 models},\ }\href@noop {} {\bibfield  {journal} {\bibinfo  {journal} {Nonlinearity}\ }\textbf {\bibinfo {volume} {19}},\ \bibinfo {pages} {217} (\bibinfo {year} {2005})}\BibitemShut {NoStop}%
\bibitem [{\citenamefont {Barashenkov}\ \emph {et~al.}(2005)\citenamefont {Barashenkov}, \citenamefont {Oxtoby},\ and\ \citenamefont {Pelinovsky}}]{barashenkov2005translationally}%
  \BibitemOpen
  \bibfield  {author} {\bibinfo {author} {\bibfnamefont {I.}~\bibnamefont {Barashenkov}}, \bibinfo {author} {\bibfnamefont {O.}~\bibnamefont {Oxtoby}},\ and\ \bibinfo {author} {\bibfnamefont {D.~E.}\ \bibnamefont {Pelinovsky}},\ }\bibfield  {title} {\bibinfo {title} {Translationally invariant discrete kinks from one-dimensional maps},\ }\href@noop {} {\bibfield  {journal} {\bibinfo  {journal} {Physical Review E—Statistical, Nonlinear, and Soft Matter Physics}\ }\textbf {\bibinfo {volume} {72}},\ \bibinfo {pages} {035602} (\bibinfo {year} {2005})}\BibitemShut {NoStop}%
\bibitem [{\citenamefont {Dmitriev}\ \emph {et~al.}(2006{\natexlab{a}})\citenamefont {Dmitriev}, \citenamefont {Kevrekidis},\ and\ \citenamefont {Yoshikawa}}]{dmitriev2006standard}%
  \BibitemOpen
  \bibfield  {author} {\bibinfo {author} {\bibfnamefont {S.}~\bibnamefont {Dmitriev}}, \bibinfo {author} {\bibfnamefont {P.}~\bibnamefont {Kevrekidis}},\ and\ \bibinfo {author} {\bibfnamefont {N.}~\bibnamefont {Yoshikawa}},\ }\bibfield  {title} {\bibinfo {title} {Standard nearest-neighbour discretizations of klein--gordon models cannot preserve both energy and linear momentum},\ }\href@noop {} {\bibfield  {journal} {\bibinfo  {journal} {Journal of Physics A: Mathematical and General}\ }\textbf {\bibinfo {volume} {39}},\ \bibinfo {pages} {7217} (\bibinfo {year} {2006}{\natexlab{a}})}\BibitemShut {NoStop}%
\bibitem [{\citenamefont {Dmitriev}\ \emph {et~al.}(2006{\natexlab{b}})\citenamefont {Dmitriev}, \citenamefont {Kevrekidis}, \citenamefont {Yoshikawa},\ and\ \citenamefont {Frantzeskakis}}]{dmitriev2006exact}%
  \BibitemOpen
  \bibfield  {author} {\bibinfo {author} {\bibfnamefont {S.}~\bibnamefont {Dmitriev}}, \bibinfo {author} {\bibfnamefont {P.}~\bibnamefont {Kevrekidis}}, \bibinfo {author} {\bibfnamefont {N.}~\bibnamefont {Yoshikawa}},\ and\ \bibinfo {author} {\bibfnamefont {D.}~\bibnamefont {Frantzeskakis}},\ }\bibfield  {title} {\bibinfo {title} {Exact static solutions for discrete $\phi$ 4 models free of the peierls-nabarro barrier: Discretized first-integral approach},\ }\href@noop {} {\bibfield  {journal} {\bibinfo  {journal} {Physical Review E—Statistical, Nonlinear, and Soft Matter Physics}\ }\textbf {\bibinfo {volume} {74}},\ \bibinfo {pages} {046609} (\bibinfo {year} {2006}{\natexlab{b}})}\BibitemShut {NoStop}%
\bibitem [{\citenamefont {Saadatmand}\ \emph {et~al.}(2024)\citenamefont {Saadatmand}, \citenamefont {Marjaneh}, \citenamefont {Askari},\ and\ \citenamefont {Weigel}}]{saadatmand2024phonons}%
  \BibitemOpen
  \bibfield  {author} {\bibinfo {author} {\bibfnamefont {D.}~\bibnamefont {Saadatmand}}, \bibinfo {author} {\bibfnamefont {A.~M.}\ \bibnamefont {Marjaneh}}, \bibinfo {author} {\bibfnamefont {A.}~\bibnamefont {Askari}},\ and\ \bibinfo {author} {\bibfnamefont {H.}~\bibnamefont {Weigel}},\ }\bibfield  {title} {\bibinfo {title} {Phonons scattering off discrete asymmetric solitons in the absence of a peierls--nabarro potential},\ }\href@noop {} {\bibfield  {journal} {\bibinfo  {journal} {Chaos, Solitons \& Fractals}\ }\textbf {\bibinfo {volume} {180}},\ \bibinfo {pages} {114550} (\bibinfo {year} {2024})}\BibitemShut {NoStop}%
\bibitem [{\citenamefont {Scharf}\ \emph {et~al.}(1992)\citenamefont {Scharf}, \citenamefont {Kivshar}, \citenamefont {S{\'a}nchez},\ and\ \citenamefont {Bishop}}]{scharf1992sine}%
  \BibitemOpen
  \bibfield  {author} {\bibinfo {author} {\bibfnamefont {R.}~\bibnamefont {Scharf}}, \bibinfo {author} {\bibfnamefont {Y.~S.}\ \bibnamefont {Kivshar}}, \bibinfo {author} {\bibfnamefont {A.}~\bibnamefont {S{\'a}nchez}},\ and\ \bibinfo {author} {\bibfnamefont {A.~R.}\ \bibnamefont {Bishop}},\ }\bibfield  {title} {\bibinfo {title} {Sine-gordon kink-antikink generation on spatially periodic potentials},\ }\href@noop {} {\bibfield  {journal} {\bibinfo  {journal} {Physical Review A}\ }\textbf {\bibinfo {volume} {45}},\ \bibinfo {pages} {R5369} (\bibinfo {year} {1992})}\BibitemShut {NoStop}%
\bibitem [{\citenamefont {Mohammadi}\ and\ \citenamefont {Dehghani}(2021)}]{mohammadi2021kink}%
  \BibitemOpen
  \bibfield  {author} {\bibinfo {author} {\bibfnamefont {M.}~\bibnamefont {Mohammadi}}\ and\ \bibinfo {author} {\bibfnamefont {R.}~\bibnamefont {Dehghani}},\ }\bibfield  {title} {\bibinfo {title} {Kink-antikink collisions in the periodic $\varphi$4 model},\ }\href@noop {} {\bibfield  {journal} {\bibinfo  {journal} {Communications in Nonlinear Science and Numerical Simulation}\ }\textbf {\bibinfo {volume} {94}},\ \bibinfo {pages} {105575} (\bibinfo {year} {2021})}\BibitemShut {NoStop}%
\bibitem [{\citenamefont {Simas}\ and\ \citenamefont {da~Hora}(2024)}]{simas2024generation}%
  \BibitemOpen
  \bibfield  {author} {\bibinfo {author} {\bibfnamefont {F.~C.}\ \bibnamefont {Simas}}\ and\ \bibinfo {author} {\bibfnamefont {E.}~\bibnamefont {da~Hora}},\ }\bibfield  {title} {\bibinfo {title} {Generation of kink-antikink pairs through the excitation of an oscillon in the $\phi^4$ model},\ }\href@noop {} {\bibfield  {journal} {\bibinfo  {journal} {arXiv preprint arXiv:2404.17848}\ } (\bibinfo {year} {2024})}\BibitemShut {NoStop}%
\bibitem [{\citenamefont {Woodhouse}\ \emph {et~al.}(2018)\citenamefont {Woodhouse}, \citenamefont {Ronellenfitsch},\ and\ \citenamefont {Dunkel}}]{woodhouse2018autonomous}%
  \BibitemOpen
  \bibfield  {author} {\bibinfo {author} {\bibfnamefont {F.~G.}\ \bibnamefont {Woodhouse}}, \bibinfo {author} {\bibfnamefont {H.}~\bibnamefont {Ronellenfitsch}},\ and\ \bibinfo {author} {\bibfnamefont {J.}~\bibnamefont {Dunkel}},\ }\bibfield  {title} {\bibinfo {title} {Autonomous actuation of zero modes in mechanical networks far from equilibrium},\ }\href@noop {} {\bibfield  {journal} {\bibinfo  {journal} {Physical Review Letters}\ }\textbf {\bibinfo {volume} {121}},\ \bibinfo {pages} {178001} (\bibinfo {year} {2018})}\BibitemShut {NoStop}%
\bibitem [{\citenamefont {Ghatak}\ \emph {et~al.}(2020)\citenamefont {Ghatak}, \citenamefont {Brandenbourger}, \citenamefont {Van~Wezel},\ and\ \citenamefont {Coulais}}]{ghatak2020observation}%
  \BibitemOpen
  \bibfield  {author} {\bibinfo {author} {\bibfnamefont {A.}~\bibnamefont {Ghatak}}, \bibinfo {author} {\bibfnamefont {M.}~\bibnamefont {Brandenbourger}}, \bibinfo {author} {\bibfnamefont {J.}~\bibnamefont {Van~Wezel}},\ and\ \bibinfo {author} {\bibfnamefont {C.}~\bibnamefont {Coulais}},\ }\bibfield  {title} {\bibinfo {title} {Observation of non-hermitian topology and its bulk--edge correspondence in an active mechanical metamaterial},\ }\href@noop {} {\bibfield  {journal} {\bibinfo  {journal} {Proceedings of the National Academy of Sciences}\ }\textbf {\bibinfo {volume} {117}},\ \bibinfo {pages} {29561} (\bibinfo {year} {2020})}\BibitemShut {NoStop}%
\bibitem [{\citenamefont {Veenstra}\ \emph {et~al.}(2024)\citenamefont {Veenstra}, \citenamefont {Gamayun}, \citenamefont {Guo}, \citenamefont {Sarvi}, \citenamefont {Meinersen},\ and\ \citenamefont {Coulais}}]{veenstra2024non}%
  \BibitemOpen
  \bibfield  {author} {\bibinfo {author} {\bibfnamefont {J.}~\bibnamefont {Veenstra}}, \bibinfo {author} {\bibfnamefont {O.}~\bibnamefont {Gamayun}}, \bibinfo {author} {\bibfnamefont {X.}~\bibnamefont {Guo}}, \bibinfo {author} {\bibfnamefont {A.}~\bibnamefont {Sarvi}}, \bibinfo {author} {\bibfnamefont {C.~V.}\ \bibnamefont {Meinersen}},\ and\ \bibinfo {author} {\bibfnamefont {C.}~\bibnamefont {Coulais}},\ }\bibfield  {title} {\bibinfo {title} {Non-reciprocal topological solitons in active metamaterials},\ }\href@noop {} {\bibfield  {journal} {\bibinfo  {journal} {Nature}\ }\textbf {\bibinfo {volume} {627}},\ \bibinfo {pages} {528} (\bibinfo {year} {2024})}\BibitemShut {NoStop}%
\bibitem [{\citenamefont {Hobart}(1965)}]{hobart1965peierls}%
  \BibitemOpen
  \bibfield  {author} {\bibinfo {author} {\bibfnamefont {R.}~\bibnamefont {Hobart}},\ }\bibfield  {title} {\bibinfo {title} {Peierls stress dependence on dislocation width},\ }\href@noop {} {\bibfield  {journal} {\bibinfo  {journal} {Journal of Applied Physics}\ }\textbf {\bibinfo {volume} {36}},\ \bibinfo {pages} {1944} (\bibinfo {year} {1965})}\BibitemShut {NoStop}%
\bibitem [{\citenamefont {Nabarro}(1989)}]{nabarro1989peierls}%
  \BibitemOpen
  \bibfield  {author} {\bibinfo {author} {\bibfnamefont {F.}~\bibnamefont {Nabarro}},\ }\bibfield  {title} {\bibinfo {title} {The peierls stress for a wide dislocation},\ }\href@noop {} {\bibfield  {journal} {\bibinfo  {journal} {Materials Science and Engineering: A}\ }\textbf {\bibinfo {volume} {113}},\ \bibinfo {pages} {315} (\bibinfo {year} {1989})}\BibitemShut {NoStop}%
\bibitem [{\citenamefont {Kane}\ and\ \citenamefont {Lubensky}(2014)}]{kane2014topological}%
  \BibitemOpen
  \bibfield  {author} {\bibinfo {author} {\bibfnamefont {C.~L.}\ \bibnamefont {Kane}}\ and\ \bibinfo {author} {\bibfnamefont {T.~C.}\ \bibnamefont {Lubensky}},\ }\bibfield  {title} {\bibinfo {title} {Topological boundary modes in isostatic lattices},\ }\href@noop {} {\bibfield  {journal} {\bibinfo  {journal} {Nature Physics}\ }\textbf {\bibinfo {volume} {10}},\ \bibinfo {pages} {39} (\bibinfo {year} {2014})}\BibitemShut {NoStop}%
\bibitem [{\citenamefont {Kevrekidis}\ and\ \citenamefont {Weinstein}(2000)}]{kevrekidis2000dynamics}%
  \BibitemOpen
  \bibfield  {author} {\bibinfo {author} {\bibfnamefont {P.}~\bibnamefont {Kevrekidis}}\ and\ \bibinfo {author} {\bibfnamefont {M.}~\bibnamefont {Weinstein}},\ }\bibfield  {title} {\bibinfo {title} {Dynamics of lattice kinks},\ }\href@noop {} {\bibfield  {journal} {\bibinfo  {journal} {Physica D: Nonlinear Phenomena}\ }\textbf {\bibinfo {volume} {142}},\ \bibinfo {pages} {113} (\bibinfo {year} {2000})}\BibitemShut {NoStop}%
\bibitem [{\citenamefont {Peyrard}\ and\ \citenamefont {Campbell}(1983)}]{peyrard1983kink}%
  \BibitemOpen
  \bibfield  {author} {\bibinfo {author} {\bibfnamefont {M.}~\bibnamefont {Peyrard}}\ and\ \bibinfo {author} {\bibfnamefont {D.~K.}\ \bibnamefont {Campbell}},\ }\bibfield  {title} {\bibinfo {title} {Kink-antikink interactions in a modified sine-gordon model},\ }\href@noop {} {\bibfield  {journal} {\bibinfo  {journal} {Physica D: Nonlinear Phenomena}\ }\textbf {\bibinfo {volume} {9}},\ \bibinfo {pages} {33} (\bibinfo {year} {1983})}\BibitemShut {NoStop}%
\bibitem [{\citenamefont {Kivshar}\ \emph {et~al.}(1991)\citenamefont {Kivshar}, \citenamefont {Fei},\ and\ \citenamefont {V{\'a}zquez}}]{kivshar1991resonant}%
  \BibitemOpen
  \bibfield  {author} {\bibinfo {author} {\bibfnamefont {Y.~S.}\ \bibnamefont {Kivshar}}, \bibinfo {author} {\bibfnamefont {Z.}~\bibnamefont {Fei}},\ and\ \bibinfo {author} {\bibfnamefont {L.}~\bibnamefont {V{\'a}zquez}},\ }\bibfield  {title} {\bibinfo {title} {Resonant soliton-impurity interactions},\ }\href@noop {} {\bibfield  {journal} {\bibinfo  {journal} {Physical review letters}\ }\textbf {\bibinfo {volume} {67}},\ \bibinfo {pages} {1177} (\bibinfo {year} {1991})}\BibitemShut {NoStop}%
\bibitem [{\citenamefont {Fei}\ \emph {et~al.}(1992)\citenamefont {Fei}, \citenamefont {Kivshar},\ and\ \citenamefont {V{\'a}zquez}}]{fei1992resonant}%
  \BibitemOpen
  \bibfield  {author} {\bibinfo {author} {\bibfnamefont {Z.}~\bibnamefont {Fei}}, \bibinfo {author} {\bibfnamefont {Y.~S.}\ \bibnamefont {Kivshar}},\ and\ \bibinfo {author} {\bibfnamefont {L.}~\bibnamefont {V{\'a}zquez}},\ }\bibfield  {title} {\bibinfo {title} {Resonant kink-impurity interactions in the $\varphi$ 4 model},\ }\href@noop {} {\bibfield  {journal} {\bibinfo  {journal} {Physical Review A}\ }\textbf {\bibinfo {volume} {46}},\ \bibinfo {pages} {5214} (\bibinfo {year} {1992})}\BibitemShut {NoStop}%
\bibitem [{\citenamefont {Vardeny}\ \emph {et~al.}(1986)\citenamefont {Vardeny}, \citenamefont {Ehrenfreund}, \citenamefont {Brafman}, \citenamefont {Horovitz}, \citenamefont {Fujimoto}, \citenamefont {Tanaka},\ and\ \citenamefont {Tanaka}}]{vardeny1986detection}%
  \BibitemOpen
  \bibfield  {author} {\bibinfo {author} {\bibfnamefont {Z.}~\bibnamefont {Vardeny}}, \bibinfo {author} {\bibfnamefont {E.}~\bibnamefont {Ehrenfreund}}, \bibinfo {author} {\bibfnamefont {O.}~\bibnamefont {Brafman}}, \bibinfo {author} {\bibfnamefont {B.}~\bibnamefont {Horovitz}}, \bibinfo {author} {\bibfnamefont {H.}~\bibnamefont {Fujimoto}}, \bibinfo {author} {\bibfnamefont {J.}~\bibnamefont {Tanaka}},\ and\ \bibinfo {author} {\bibfnamefont {M.}~\bibnamefont {Tanaka}},\ }\bibfield  {title} {\bibinfo {title} {Detection of soliton shape modes in polyacetylene},\ }\href@noop {} {\bibfield  {journal} {\bibinfo  {journal} {Physical review letters}\ }\textbf {\bibinfo {volume} {57}},\ \bibinfo {pages} {2995} (\bibinfo {year} {1986})}\BibitemShut {NoStop}%
\bibitem [{\citenamefont {Mielenz}\ \emph {et~al.}(2013)\citenamefont {Mielenz}, \citenamefont {Brox}, \citenamefont {Kahra}, \citenamefont {Leschhorn}, \citenamefont {Albert}, \citenamefont {Sch{\"a}tz}, \citenamefont {Landa},\ and\ \citenamefont {Reznik}}]{mielenz2013trapping}%
  \BibitemOpen
  \bibfield  {author} {\bibinfo {author} {\bibfnamefont {M.}~\bibnamefont {Mielenz}}, \bibinfo {author} {\bibfnamefont {J.}~\bibnamefont {Brox}}, \bibinfo {author} {\bibfnamefont {S.}~\bibnamefont {Kahra}}, \bibinfo {author} {\bibfnamefont {G.}~\bibnamefont {Leschhorn}}, \bibinfo {author} {\bibfnamefont {M.}~\bibnamefont {Albert}}, \bibinfo {author} {\bibfnamefont {T.}~\bibnamefont {Sch{\"a}tz}}, \bibinfo {author} {\bibfnamefont {H.}~\bibnamefont {Landa}},\ and\ \bibinfo {author} {\bibfnamefont {B.}~\bibnamefont {Reznik}},\ }\bibfield  {title} {\bibinfo {title} {Trapping of topological-structural defects in coulomb crystals},\ }\href@noop {} {\bibfield  {journal} {\bibinfo  {journal} {Physical review letters}\ }\textbf {\bibinfo {volume} {110}},\ \bibinfo {pages} {133004} (\bibinfo {year} {2013})}\BibitemShut {NoStop}%
\bibitem [{\citenamefont {Brox}\ \emph {et~al.}(2017)\citenamefont {Brox}, \citenamefont {Kiefer}, \citenamefont {Bujak}, \citenamefont {Schaetz},\ and\ \citenamefont {Landa}}]{brox2017spectroscopy}%
  \BibitemOpen
  \bibfield  {author} {\bibinfo {author} {\bibfnamefont {J.}~\bibnamefont {Brox}}, \bibinfo {author} {\bibfnamefont {P.}~\bibnamefont {Kiefer}}, \bibinfo {author} {\bibfnamefont {M.}~\bibnamefont {Bujak}}, \bibinfo {author} {\bibfnamefont {T.}~\bibnamefont {Schaetz}},\ and\ \bibinfo {author} {\bibfnamefont {H.}~\bibnamefont {Landa}},\ }\bibfield  {title} {\bibinfo {title} {Spectroscopy and directed transport of topological solitons in crystals of trapped ions},\ }\href@noop {} {\bibfield  {journal} {\bibinfo  {journal} {Physical review letters}\ }\textbf {\bibinfo {volume} {119}},\ \bibinfo {pages} {153602} (\bibinfo {year} {2017})}\BibitemShut {NoStop}%
\bibitem [{\citenamefont {Ni}\ \emph {et~al.}(2023)\citenamefont {Ni}, \citenamefont {Yves}, \citenamefont {Krasnok},\ and\ \citenamefont {Alu}}]{ni2023topological}%
  \BibitemOpen
  \bibfield  {author} {\bibinfo {author} {\bibfnamefont {X.}~\bibnamefont {Ni}}, \bibinfo {author} {\bibfnamefont {S.}~\bibnamefont {Yves}}, \bibinfo {author} {\bibfnamefont {A.}~\bibnamefont {Krasnok}},\ and\ \bibinfo {author} {\bibfnamefont {A.}~\bibnamefont {Alu}},\ }\bibfield  {title} {\bibinfo {title} {Topological metamaterials},\ }\href@noop {} {\bibfield  {journal} {\bibinfo  {journal} {Chemical Reviews}\ }\textbf {\bibinfo {volume} {123}},\ \bibinfo {pages} {7585} (\bibinfo {year} {2023})}\BibitemShut {NoStop}%
\bibitem [{\citenamefont {Rocklin}\ \emph {et~al.}(2017)\citenamefont {Rocklin}, \citenamefont {Zhou}, \citenamefont {Sun},\ and\ \citenamefont {Mao}}]{rocklin2017transformable}%
  \BibitemOpen
  \bibfield  {author} {\bibinfo {author} {\bibfnamefont {D.~Z.}\ \bibnamefont {Rocklin}}, \bibinfo {author} {\bibfnamefont {S.}~\bibnamefont {Zhou}}, \bibinfo {author} {\bibfnamefont {K.}~\bibnamefont {Sun}},\ and\ \bibinfo {author} {\bibfnamefont {X.}~\bibnamefont {Mao}},\ }\bibfield  {title} {\bibinfo {title} {Transformable topological mechanical metamaterials},\ }\href@noop {} {\bibfield  {journal} {\bibinfo  {journal} {Nature communications}\ }\textbf {\bibinfo {volume} {8}},\ \bibinfo {pages} {14201} (\bibinfo {year} {2017})}\BibitemShut {NoStop}%
\bibitem [{\citenamefont {Rodr{\'\i}guez}\ \emph {et~al.}(2023)\citenamefont {Rodr{\'\i}guez}, \citenamefont {Calius}, \citenamefont {Khatibi}, \citenamefont {Orifici},\ and\ \citenamefont {Das}}]{rodriguez2023mechanical}%
  \BibitemOpen
  \bibfield  {author} {\bibinfo {author} {\bibfnamefont {S.~E.}\ \bibnamefont {Rodr{\'\i}guez}}, \bibinfo {author} {\bibfnamefont {E.}~\bibnamefont {Calius}}, \bibinfo {author} {\bibfnamefont {A.}~\bibnamefont {Khatibi}}, \bibinfo {author} {\bibfnamefont {A.}~\bibnamefont {Orifici}},\ and\ \bibinfo {author} {\bibfnamefont {R.}~\bibnamefont {Das}},\ }\bibfield  {title} {\bibinfo {title} {Mechanical metamaterial systems as transformation mechanisms},\ }\href@noop {} {\bibfield  {journal} {\bibinfo  {journal} {Extreme Mechanics Letters}\ }\textbf {\bibinfo {volume} {61}},\ \bibinfo {pages} {101985} (\bibinfo {year} {2023})}\BibitemShut {NoStop}%
\bibitem [{\citenamefont {Peplow}(2015)}]{peplow2015tiniest}%
  \BibitemOpen
  \bibfield  {author} {\bibinfo {author} {\bibfnamefont {M.}~\bibnamefont {Peplow}},\ }\bibfield  {title} {\bibinfo {title} {The tiniest lego: a tale of nanoscale motors, rotors, switches and pumps.},\ }\href@noop {} {\bibfield  {journal} {\bibinfo  {journal} {Nature}\ }\textbf {\bibinfo {volume} {525}} (\bibinfo {year} {2015})}\BibitemShut {NoStop}%
\bibitem [{\citenamefont {Nirody}\ \emph {et~al.}(2017)\citenamefont {Nirody}, \citenamefont {Sun},\ and\ \citenamefont {Lo}}]{nirody2017biophysicist}%
  \BibitemOpen
  \bibfield  {author} {\bibinfo {author} {\bibfnamefont {J.~A.}\ \bibnamefont {Nirody}}, \bibinfo {author} {\bibfnamefont {Y.-R.}\ \bibnamefont {Sun}},\ and\ \bibinfo {author} {\bibfnamefont {C.-J.}\ \bibnamefont {Lo}},\ }\bibfield  {title} {\bibinfo {title} {The biophysicist’s guide to the bacterial flagellar motor},\ }\href@noop {} {\bibfield  {journal} {\bibinfo  {journal} {Advances in Physics: X}\ }\textbf {\bibinfo {volume} {2}},\ \bibinfo {pages} {324} (\bibinfo {year} {2017})}\BibitemShut {NoStop}%
\bibitem [{\citenamefont {Singh}\ \emph {et~al.}(2024)\citenamefont {Singh}, \citenamefont {Sharma}, \citenamefont {Afanzar}, \citenamefont {Goldfarb}, \citenamefont {Maklashina}, \citenamefont {Eisenbach}, \citenamefont {Cecchini},\ and\ \citenamefont {Iverson}}]{singh2024cryoem}%
  \BibitemOpen
  \bibfield  {author} {\bibinfo {author} {\bibfnamefont {P.~K.}\ \bibnamefont {Singh}}, \bibinfo {author} {\bibfnamefont {P.}~\bibnamefont {Sharma}}, \bibinfo {author} {\bibfnamefont {O.}~\bibnamefont {Afanzar}}, \bibinfo {author} {\bibfnamefont {M.~H.}\ \bibnamefont {Goldfarb}}, \bibinfo {author} {\bibfnamefont {E.}~\bibnamefont {Maklashina}}, \bibinfo {author} {\bibfnamefont {M.}~\bibnamefont {Eisenbach}}, \bibinfo {author} {\bibfnamefont {G.}~\bibnamefont {Cecchini}},\ and\ \bibinfo {author} {\bibfnamefont {T.}~\bibnamefont {Iverson}},\ }\bibfield  {title} {\bibinfo {title} {Cryoem structures reveal how the bacterial flagellum rotates and switches direction},\ }\href@noop {} {\bibfield  {journal} {\bibinfo  {journal} {Nature Microbiology}\ ,\ \bibinfo {pages} {1}} (\bibinfo {year} {2024})}\BibitemShut {NoStop}%
\bibitem [{\citenamefont {Shi}\ \emph {et~al.}(2022)\citenamefont {Shi}, \citenamefont {Pumm}, \citenamefont {Isensee}, \citenamefont {Zhao}, \citenamefont {Verschueren}, \citenamefont {Martin-Gonzalez}, \citenamefont {Golestanian}, \citenamefont {Dietz},\ and\ \citenamefont {Dekker}}]{shi2022sustained}%
  \BibitemOpen
  \bibfield  {author} {\bibinfo {author} {\bibfnamefont {X.}~\bibnamefont {Shi}}, \bibinfo {author} {\bibfnamefont {A.-K.}\ \bibnamefont {Pumm}}, \bibinfo {author} {\bibfnamefont {J.}~\bibnamefont {Isensee}}, \bibinfo {author} {\bibfnamefont {W.}~\bibnamefont {Zhao}}, \bibinfo {author} {\bibfnamefont {D.}~\bibnamefont {Verschueren}}, \bibinfo {author} {\bibfnamefont {A.}~\bibnamefont {Martin-Gonzalez}}, \bibinfo {author} {\bibfnamefont {R.}~\bibnamefont {Golestanian}}, \bibinfo {author} {\bibfnamefont {H.}~\bibnamefont {Dietz}},\ and\ \bibinfo {author} {\bibfnamefont {C.}~\bibnamefont {Dekker}},\ }\bibfield  {title} {\bibinfo {title} {Sustained unidirectional rotation of a self-organized dna rotor on a nanopore},\ }\href@noop {} {\bibfield  {journal} {\bibinfo  {journal} {Nature Physics}\ }\textbf {\bibinfo {volume} {18}},\ \bibinfo {pages} {1105} (\bibinfo {year} {2022})}\BibitemShut {NoStop}%
\bibitem [{\citenamefont {Kim}\ \emph {et~al.}(2014)\citenamefont {Kim}, \citenamefont {Xu}, \citenamefont {Guo},\ and\ \citenamefont {Fan}}]{kim2014ultrahigh}%
  \BibitemOpen
  \bibfield  {author} {\bibinfo {author} {\bibfnamefont {K.}~\bibnamefont {Kim}}, \bibinfo {author} {\bibfnamefont {X.}~\bibnamefont {Xu}}, \bibinfo {author} {\bibfnamefont {J.}~\bibnamefont {Guo}},\ and\ \bibinfo {author} {\bibfnamefont {D.}~\bibnamefont {Fan}},\ }\bibfield  {title} {\bibinfo {title} {Ultrahigh-speed rotating nanoelectromechanical system devices assembled from nanoscale building blocks},\ }\href@noop {} {\bibfield  {journal} {\bibinfo  {journal} {Nature communications}\ }\textbf {\bibinfo {volume} {5}},\ \bibinfo {pages} {3632} (\bibinfo {year} {2014})}\BibitemShut {NoStop}%
\bibitem [{\citenamefont {Manton}\ and\ \citenamefont {Sutcliffe}(2004)}]{manton2004topological}%
  \BibitemOpen
  \bibfield  {author} {\bibinfo {author} {\bibfnamefont {N.}~\bibnamefont {Manton}}\ and\ \bibinfo {author} {\bibfnamefont {P.}~\bibnamefont {Sutcliffe}},\ }\href@noop {} {\emph {\bibinfo {title} {Topological solitons}}}\ (\bibinfo  {publisher} {Cambridge University Press},\ \bibinfo {year} {2004})\BibitemShut {NoStop}%
\bibitem [{\citenamefont {Gilat}\ and\ \citenamefont {Subramaniam}(2013)}]{gilat2013numerical}%
  \BibitemOpen
  \bibfield  {author} {\bibinfo {author} {\bibfnamefont {A.}~\bibnamefont {Gilat}}\ and\ \bibinfo {author} {\bibfnamefont {V.}~\bibnamefont {Subramaniam}},\ }\bibfield  {title} {\bibinfo {title} {Numerical methods for engineers and scientists},\ }\href@noop {} {\bibfield  {journal} {\bibinfo  {journal} {An Introduction with Applications Using MATLAB,}\ }\textbf {\bibinfo {volume} {20014}} (\bibinfo {year} {2013})}\BibitemShut {NoStop}%
\bibitem [{\citenamefont {Calladine}(1978)}]{calladine1978buckminster}%
  \BibitemOpen
  \bibfield  {author} {\bibinfo {author} {\bibfnamefont {C.~R.}\ \bibnamefont {Calladine}},\ }\bibfield  {title} {\bibinfo {title} {Buckminster fuller's “tensegrity” structures and clerk maxwell's rules for the construction of stiff frames},\ }\href@noop {} {\bibfield  {journal} {\bibinfo  {journal} {International journal of solids and structures}\ }\textbf {\bibinfo {volume} {14}},\ \bibinfo {pages} {161} (\bibinfo {year} {1978})}\BibitemShut {NoStop}%
\bibitem [{\citenamefont {Sun}\ \emph {et~al.}(2012)\citenamefont {Sun}, \citenamefont {Souslov}, \citenamefont {Mao},\ and\ \citenamefont {Lubensky}}]{sun2012surface}%
  \BibitemOpen
  \bibfield  {author} {\bibinfo {author} {\bibfnamefont {K.}~\bibnamefont {Sun}}, \bibinfo {author} {\bibfnamefont {A.}~\bibnamefont {Souslov}}, \bibinfo {author} {\bibfnamefont {X.}~\bibnamefont {Mao}},\ and\ \bibinfo {author} {\bibfnamefont {T.~C.}\ \bibnamefont {Lubensky}},\ }\bibfield  {title} {\bibinfo {title} {Surface phonons, elastic response, and conformal invariance in twisted kagome lattices},\ }\href@noop {} {\bibfield  {journal} {\bibinfo  {journal} {Proceedings of the National Academy of Sciences}\ }\textbf {\bibinfo {volume} {109}},\ \bibinfo {pages} {12369} (\bibinfo {year} {2012})}\BibitemShut {NoStop}%
\bibitem [{\citenamefont {Paulose}\ \emph {et~al.}(2015)\citenamefont {Paulose}, \citenamefont {Chen},\ and\ \citenamefont {Vitelli}}]{paulose2015topological}%
  \BibitemOpen
  \bibfield  {author} {\bibinfo {author} {\bibfnamefont {J.}~\bibnamefont {Paulose}}, \bibinfo {author} {\bibfnamefont {B.~G.-g.}\ \bibnamefont {Chen}},\ and\ \bibinfo {author} {\bibfnamefont {V.}~\bibnamefont {Vitelli}},\ }\bibfield  {title} {\bibinfo {title} {Topological modes bound to dislocations in mechanical metamaterials},\ }\href@noop {} {\bibfield  {journal} {\bibinfo  {journal} {Nature Physics}\ }\textbf {\bibinfo {volume} {11}},\ \bibinfo {pages} {153} (\bibinfo {year} {2015})}\BibitemShut {NoStop}%
\bibitem [{\citenamefont {Sato}\ and\ \citenamefont {Tanaka}(2018)}]{sato2018solitons}%
  \BibitemOpen
  \bibfield  {author} {\bibinfo {author} {\bibfnamefont {K.}~\bibnamefont {Sato}}\ and\ \bibinfo {author} {\bibfnamefont {R.}~\bibnamefont {Tanaka}},\ }\bibfield  {title} {\bibinfo {title} {Solitons in one-dimensional mechanical linkage},\ }\href@noop {} {\bibfield  {journal} {\bibinfo  {journal} {Physical Review E}\ }\textbf {\bibinfo {volume} {98}},\ \bibinfo {pages} {013001} (\bibinfo {year} {2018})}\BibitemShut {NoStop}%
\bibitem [{\citenamefont {Khomeriki}\ and\ \citenamefont {Leon}(2008)}]{khomeriki2008tristability}%
  \BibitemOpen
  \bibfield  {author} {\bibinfo {author} {\bibfnamefont {R.}~\bibnamefont {Khomeriki}}\ and\ \bibinfo {author} {\bibfnamefont {J.}~\bibnamefont {Leon}},\ }\bibfield  {title} {\bibinfo {title} {Tristability in the pendula chain},\ }\href@noop {} {\bibfield  {journal} {\bibinfo  {journal} {Physical Review E—Statistical, Nonlinear, and Soft Matter Physics}\ }\textbf {\bibinfo {volume} {78}},\ \bibinfo {pages} {057202} (\bibinfo {year} {2008})}\BibitemShut {NoStop}%
\bibitem [{\citenamefont {Wu}\ \emph {et~al.}(2018)\citenamefont {Wu}, \citenamefont {Zheng},\ and\ \citenamefont {Wang}}]{wu2018metastable}%
  \BibitemOpen
  \bibfield  {author} {\bibinfo {author} {\bibfnamefont {Z.}~\bibnamefont {Wu}}, \bibinfo {author} {\bibfnamefont {Y.}~\bibnamefont {Zheng}},\ and\ \bibinfo {author} {\bibfnamefont {K.}~\bibnamefont {Wang}},\ }\bibfield  {title} {\bibinfo {title} {Metastable modular metastructures for on-demand reconfiguration of band structures and nonreciprocal wave propagation},\ }\href@noop {} {\bibfield  {journal} {\bibinfo  {journal} {Physical Review E}\ }\textbf {\bibinfo {volume} {97}},\ \bibinfo {pages} {022209} (\bibinfo {year} {2018})}\BibitemShut {NoStop}%
\bibitem [{\citenamefont {Hwang}\ and\ \citenamefont {Arrieta}(2021)}]{hwang2021extreme}%
  \BibitemOpen
  \bibfield  {author} {\bibinfo {author} {\bibfnamefont {M.}~\bibnamefont {Hwang}}\ and\ \bibinfo {author} {\bibfnamefont {A.~F.}\ \bibnamefont {Arrieta}},\ }\bibfield  {title} {\bibinfo {title} {Extreme frequency conversion from soliton resonant interactions},\ }\href@noop {} {\bibfield  {journal} {\bibinfo  {journal} {Physical Review Letters}\ }\textbf {\bibinfo {volume} {126}},\ \bibinfo {pages} {073902} (\bibinfo {year} {2021})}\BibitemShut {NoStop}%
\end{thebibliography}
%

\end{document}